%% file: main_ttor.tex
\documentclass[english,conference,10pt]{IEEEtran} 
\IEEEoverridecommandlockouts
\usepackage{etoolbox}
\usepackage{amsmath,amssymb,amsfonts,amsopn,mathrsfs,calligra}
\usepackage{graphicx,subfig,algorithm}
\usepackage[noend]{algpseudocode}
\usepackage{array,multirow}
\usepackage{tikz,pgfplots}
\usetikzlibrary{patterns}
\usetikzlibrary{shapes}
\usetikzlibrary{quotes,arrows.meta}
\usetikzlibrary{pgfplots.groupplots}
\pgfplotsset{compat=1.12}
\usetikzlibrary{external}
\usetikzlibrary{positioning}
\usetikzlibrary{snakes}
\usepackage{pgfplotstable}
\usepackage{hyperref,cleveref,listings,filecontents}
\usepackage[final]{changes}

\newtheorem{theorem}{Theorem}
\newtheorem{lemma}{Lemma}
\newtheorem{definition}{Definition}

\definecolor{green1}{cmyk}{1,0,1,0}
\lstdefinestyle{cstyle}{
  basicstyle=\ttfamily,
  language=C++,
  keywordstyle=\color{blue}\ttfamily,
  stringstyle=\color{red}\ttfamily,
  commentstyle=\color{green1}\ttfamily,
}
\pgfplotsset{
    discard if/.style 2 args={
        x filter/.append code={
            \edef\tempa{\thisrow{#1}}
            \edef\tempb{#2}
            \ifx\tempa\tempb
                \def\pgfmathresult{inf}
            \fi
        }
    },
    discard if not/.style 2 args={
        x filter/.append code={
            \edef\tempa{\thisrow{#1}}
            \edef\tempb{#2}
            \ifx\tempa\tempb
            \else
                \def\pgfmathresult{inf}
            \fi
        }
    }
}

\newcommand{\Ttor}{\texttt{TTor}}

\title{TaskTorrent: a Lightweight Distributed\\Task-Based Runtime System in C++
\thanks{L.C.\ was supported by a fellowship from Total S.A.\ Y.Q.\ and E.D.\ were supported by Grant no.\ 80NSSC18M0152 from NASA.}
}

\author{
    \IEEEauthorblockN{L{\'e}opold Cambier}
    \IEEEauthorblockA{ICME, Stanford University\\\lstinline{lcambier@stanford.edu}}
    \and 
    \IEEEauthorblockN{Yizhou Qian} 
    \IEEEauthorblockA{ICME, Stanford University\\\lstinline{yzqian@stanford.edu}}
    \and 
    \IEEEauthorblockN{Eric Darve}
    \IEEEauthorblockA{ME \& ICME, Stanford University\\\lstinline{darve@stanford.edu}}
}

\begin{document}

\maketitle

\begin{abstract}
We present TaskTorrent, a lightweight distributed task-based runtime in C++.
TaskTorrent uses a parametrized task graph to express the task DAG, and 
one-sided active messages to trigger remote tasks asynchronously.
As a result the task DAG is completely distributed and discovered in parallel.
It is a C++14 library and only depends on MPI.
We explain the API and the implementation.
We perform a series of benchmarks against \added{StarPU and ScaLAPACK.}
Micro benchmarks show it has a minimal overhead compared to other solutions.
We then apply it to two large linear algebra problems. 
TaskTorrent scales very well to thousands of cores, exhibiting good weak and strong scalings.
\end{abstract}

\input{introduction}
\input{contributions}
\input{organization}
\input{philosophy}
\input{api}
\input{implementation}
\input{completion}
\input{benchmarks}
\input{micro_benchmarks}
\input{gemm}
\input{dense_cholesky}
\input{previous}
\input{conclusion}

\bibliography{references}
\bibliographystyle{IEEEtran}

\newpage
\appendix

\input{appendix}

\end{document}

%% file: introduction.tex
\section{Introduction}

\subsection{Parallel runtime systems}

\begin{figure}[b]
    \centering
    \subfloat[\label{fig:mpi}A typical MPI program]{
        \begin{minipage}{0.22\textwidth}
\begin{lstlisting}[language=C++,basicstyle=\footnotesize\ttfamily]
for (auto i : local0)
  compute0(i);
if ([...]) 
  { MPI_Send(m, ...); }
else
  { MPI_Recv(m, ...); }
for (auto i : local1)
  compute1(i);
\end{lstlisting}
        \end{minipage}
    } 
    \subfloat[\label{fig:dag}Example of DAG of tasks.]{
        \begin{minipage}{0.22\textwidth}
        \centering
        \begin{tikzpicture}[tips=proper]
            \node (a) at (0,0) {$a$};
            \node [below=0.5cm of a] (b) {$b$};
            \node [above right=0.1cm and 0.5cm of a] (c) {$c$};
            \node [below=0.5cm of c] (d) {$d$};
            \node [below right=0.1cm and 0.5cm of b] (e) {$e$};
            \node [right=0.5cm of c] (f) {$f$};
            \node [above right=0.1cm and 0.5cm of e] (g) {$g$};
            \draw [->,>=stealth] (a) edge (c);
            \draw [->,>=stealth] (a) edge (d);
            \draw [->,>=stealth] (b) edge (d);
            \draw [->,>=stealth] (b) edge (e);
            \draw [->,>=stealth] (c) edge (f);
            \draw [->,>=stealth] (d) edge (g);
            \draw [->,>=stealth] (e) edge (g);
        \end{tikzpicture}
        \end{minipage}
    }
    \caption{More parallelism can be extracted using a tasks DAG:
    task $d$ needs to wait for task $a$ and $b$. However, task $f$ can run as soon as task $c$ has finished.}
    \label{fig:fork_join_spmd_dag}
\end{figure}
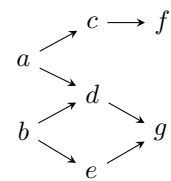

Classical parallel computing has traditionally followed a fork-join (as in OpenMP) or bulk-synchronous (MPI) approach.
(\Cref{fig:mpi} shows the skeleton of a typical MPI program).
This has many advantages, including ease of programming and predictable performance.
It has however a key downside: \added{many points of synchronization during execution are added,} even when not necessary.

Runtime systems take a different approach. 
The key concept is to express computations as a graph of tasks with dependencies between them (\Cref{fig:dag}). 
This graph is directed and acyclic, and we will later refer to it as the task DAG.
Given the DAG, the runtime system is able to extract parallelism by identifying which tasks can run in parallel.
Tasks are then assigned to processors (either individual cores, nodes, accelerators, etc).
The advantage of this method is that it removes \added{all} unnecessary synchronization points.

\subsection{Existing approaches to describe the DAG} \label{subsec:stf_ptg}

A key design choice in runtime systems is how to express the DAG.\@
At a high-level, two approaches have been primarily used.

\subsubsection{Sequential Task Flow (STF)}

In this approach, the graph is discovered by the runtime using a sequential semantics, \added{that is, typically, on each node a single thread is responsible for building the DAG.}
\added{Different mechanisms to compute task dependencies can be used.}
Often, this takes the form of inferring dependencies based on \added{specifiying data sharing rules}
(e.g., READ, WRITE, READWRITE).

This is the approach taken by Legion/Regent \cite{bauer2014legion}
\footnote{Legion is the name of the lower level C++ API, while Regent is 
the name of the higher-level language based on Lua.} and StarPU \cite{augonnet2011starpu}.
In both, the user first defines data regions and tasks operating on those regions (as inputs or outputs). 
Regent maintains a global view of the data, and \added{data regions correspond to a partitioning of the data.}
\added{The user is also able to write mappers to indicate how to map and schedule tasks to the available hardware.}
StarPU uses data handles referring to distributed memory buffers.
The program is then written in a sequential style (with for loops, if/else statements, etc.), creating tasks on previously registered data regions.
The runtime system then discovers task dependencies, builds the DAG and executes tasks \added{in parallel.}

The key in the STF approach is that the DAG has to be discovered \added{through sequential enumeration.}
\added{This restriction may have performance implications but is attractive to the programmer,} since the program is easy to write and understand.

\subsubsection{Parametrized Task Graph (PTG)}

The PTG approach is another method to express the DAG.
Using some index space (\lstinline{K}) to index all tasks, functions of \lstinline{K} are used
to express tasks and their dependencies.
As an example, \added{the DAG could be defined by specifying three functions of \lstinline{K} (other choices are possible):} one for the in-dependencies, one for the computational task itself and one for the out-dependencies. \added{By running these functions as needed, the runtime discovers the DAG dynamically.}

PaRSEC \cite{6654146} takes that approach, using a custom language (JDF) to express the PTG.\@
\added{In PaRSEC, in and out-dependencies specifications contain both tasks and data.}

The PTG format has multiple advantages.
Since task in/out-dependencies can be independently queried at any time, it simplifies task management, leading to minimal overhead during execution.
\added{It also naturally scales by parallelizing both the DAG creation and DAG execution.}
In contrast, a STF code uses, in its purest form, a single thread to discover the DAG.\@
\added{It also removes the need to store in memory large portions of the DAG of tasks.}
Instead, the runtime can query the relevant functions only as needed and discover the DAG piece by piece.

The main drawback of the PTG approach is that the program no longer has a sequential semantics, which 
makes it harder to understand the program's behavior at first sight.
\Cref{fig:stf_vs_ptg} illustrates at a higher level the differences between the STF and the PTG approach.
    
\begin{figure}
    \centering
    \subfloat[][\label{fig:stf}STF based program. 
    Dependencies are inferred through data sharing rules.
    ]{
\begin{lstlisting}[basicstyle=\footnotesize\ttfamily]
/** Define task **/
void task(...) {...}
/** Register data **/
data = [...]
/** Process DAG **/
for(k ...)
    task(data[k], 
         data[k1], 
         ...)
\end{lstlisting}
    } \;
    \subfloat[][\label{fig:ptg}PTG based program.
    Task dependencies are defined using functions over \lstinline{K}.
    Computation is triggered by seeding the initial tasks.]{
\begin{lstlisting}[basicstyle=\footnotesize\ttfamily]
/** Task deps. expressed 
 *  as functions of K 
 **/
in_deps  = (K k){...}
task     = (K k){...}
out_deps = (K k){...}
/** Seed tasks **/
for(k in kinit) 
    start(k)
\end{lstlisting}
    }
    \caption{\added{Schematic of STF and PTG programs.}}
    \label{fig:stf_vs_ptg}
\end{figure}

%% file: contributions.tex
\subsection{Contributions} \label{sec:contributions}

In this paper, we present TaskTorrent (\Ttor{}).
\Ttor{} is a lightweight, distributed task based runtime that uses a PTG approach.
Our main contributions are:
\begin{itemize}
    \item We show how to combine a PTG approach with one-sided active messages.
    \item \added{A mathematical proof is provided for the correctness of our implementation.}
    \item \added{We benchmark \Ttor{} and show that it matches or exceeds performance of StarPU on sample problems.}
\end{itemize}
\Ttor{} has a couple of notable features compared to existing solutions
\begin{itemize}
    \item It is a C++14 library with no dependencies other than MPI.
    \item \Ttor{}'s implementation leads to a small overhead and handles well small task granularity (about 10 $\mu{}s$ and up). This means that \Ttor{} can be used on any existing code, without needing to fuse or redefine tasks, or change existing algorithms.
    \item \added{Default options in \Ttor{} are designed to provide good performance ``out-of-the-box'' without requiring the user to tune or optimize internal parameters or functionalities of the library.}
    \item The user can use their own data structures without having to wrap their data in opaque data structures.
    \item \added{It is perfectly scalable in the following sense. Consider a provably scalable numerical algorithm (e.g., there exists an iso-efficiency curve). Assume that (1) the parallel computer is composed of nodes with a bounded number of cores, but with an unbounded number of nodes, and (2) that each node in the DAG has a bounded number of dependencies. Then if the algorithm is executed using \Ttor{} it will remain scalable. Said more simply, \Ttor{} does not introduce any parallel bottleneck.}
\end{itemize}
We emphasize that \Ttor{} is a general purpose runtime system. 
The applications in this paper are mostly in dense linear algebra, but there are no
features or optimizations that are specific to linear algebra in this version of \Ttor{}.

%% file: organization.tex
\subsection{Organization of the paper}

This paper is organized as follows.
\Cref{sec:features} describes \Ttor{}'s API and implementation.
\Cref{sec:benchmarks} compares \Ttor{} to StarPU and ScaLAPACK, first validating its 
shared memory component and then comparing it on large linear algebra problems.
We finally survey previous work in \Cref{sec:previous_work} before concluding.

%% file: philosophy.tex
\section{TaskTorrent}
\label{sec:features}

\Ttor{} uses a PTG.
The DAG is expressed by providing at least three functions: 
(1) one returning the number of in-dependencies of every task;
(2) one that runs the computational task and fulfills dependencies on other tasks;
(3) one returning the thread each task should be mapped to \added{(an option is provided to bound the task to the thread or leave it stealable).}
When their dependencies are satisfied, tasks are inserted into a thread pool, where a work-stealing algorithm
keeps the load balanced between the threads.

Tasks then run and fulfill other tasks' dependencies, locally (on the same rank) or remotely on a different rank.
In the case of remote dependencies, since all computations are asynchronous, the receiver rank cannot explicitly wait for data to arrive.
Hence, one-sided active messages are used. An active message (AM) is a pair (function, data). 
Once the AM arrives on the receiver, the function is run with the data passed as argument.
This is typically used to store the data and fulfill dependencies, eventually triggering more tasks.

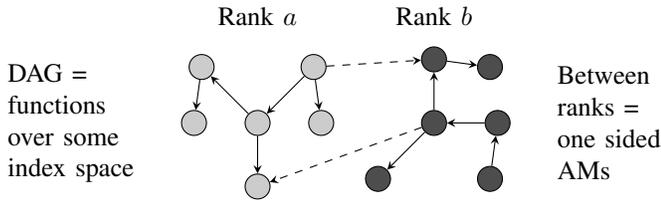
\begin{figure}
    \centering
    \begin{tikzpicture}
        \tikzstyle{every path}=[->,>=stealth];
        \def\L{0.5cm}; 

        \tikzstyle{every node}=[draw,circle,fill=black!20];
        \node (a) {};
        \node [above left = \L and \L of a] (b) {};
        \node [above right = \L and \L of a] (c) {};
        \node [left = \L of a] (d) {};
        \node [right = \L of a] (e) {};
        \node [below = \L of a] (f) {};
        
        \path (a) edge (b);
        \path (b) edge (d);
        \path (c) edge (e);
        \path (c) edge (a);
        \path (a) edge (f);
        
        \tikzstyle{every node}=[draw,circle,fill=black!70];
        \node [right = 2cm of a] (a2) {};
        \node [below left = \L and \L of a2] (b2) {};
        \node [below right = \L and \L of a2] (c2) {};
        \node [above = \L of a2] (d2) {};
        \node [right = \L of a2] (e2) {};
        \node [above right = \L and \L of a2] (f2) {};
        
        \path (c2) edge (e2);
        \path (e2) edge (a2);
        \path (a2) edge (d2);
        \path (a2) edge (b2);
        \path (d2) edge (f2);
        
        \path (a2) edge[dashed] (f);
        \path (c)  edge[dashed] (d2);
        
        \tikzstyle{every node}=[];
        \node [above = 2*\L of a] {Rank $a$};
        \node [above = 2*\L of a2] {Rank $b$};
        \node [left = \L of d,align=left] {DAG =\\functions\\over some\\index space};
        \node [right = \L of e2,align=left] {Between\\ranks =\\one sided\\AMs};
        
      \end{tikzpicture}
    \caption{The model of \Ttor{}: a distributed graph of tasks expressed using a parametrized task graph (solid arrows), with explicit active messages (dashed arrows) between ranks to asynchronously insert/trigger tasks.}
    \label{fig:ttor_distributed_dag}
\end{figure}

This approach means \Ttor{} never needs to store the full DAG.
Task dependencies are queried only when needed, and the DAG is discovered piece by piece.
In particular, \Ttor{} becomes aware of the existence of a specific task \textbf{only} when a task fulfills its first dependency. This makes \Ttor{} scalable and lightweight. The full DAG is never stored or even explored by any specific thread or rank, and the task management overhead is minimal. \Cref{fig:ttor_distributed_dag} illustrates this local DAG + AM model.

%% file: api.tex
\subsection{API Description} \label{sec:api}

\Ttor{}'s API can be divided into two parts, a shared memory component (expressing the PTG) and 
a distributed component (used for AMs). 
The combination of those two features is what distinguishes \Ttor{} from other solutions
and is one of the factors that makes \Ttor{} lightweight.


\subsubsection{Shared memory components}


\paragraph{Threadpool}
A \lstinline{Threadpool} is a fixed set of threads that receive and process tasks. 
A threadpool with \lstinline{n_threads} threads can be created by \lstinline[breaklines=true,breakatwhitespace=true]{Threadpool tp(n_threads, &comm)}.
(\lstinline{comm} is a \lstinline{Communicator}; see \Cref{sec:api_distributed}).
Tasks can be inserted directly in the threadpool, but typically this is done using a \lstinline{Taskflow}.
The threadpool joins when calling \lstinline{tp.join()}.
This returns when all the threads are idle and all communications have completed.
\Cref{sec:distributed_completion} explains in details the distributed completion mechanism.

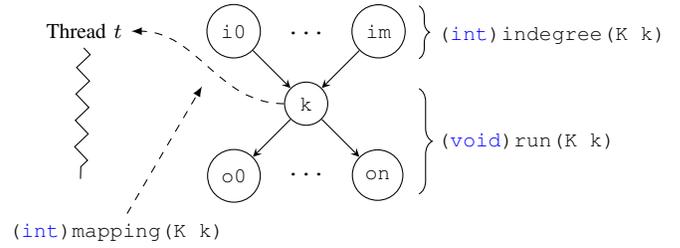
\begin{figure}
    \centering
    \begin{tikzpicture}[node distance = 0.8cm]
        \tikzstyle{every node}=[font=\footnotesize,text centered];
        \tikzset{zigzag/.style={decorate, decoration=zigzag}};

        \node [circle,draw] (task_k) at (0,0) {\lstinline{k}};
        \node [circle,draw,above left=0.5cm and 0.5cm of task_k] (k0) {\lstinline{i0}};
        \node [above=0.5cm of task_k] {\lstinline{...}};
        \node [circle,draw,above right=0.5cm and 0.5cm of task_k] (kn) {\lstinline{im}};
        
        \node [circle,draw,below left=0.5cm and 0.5cm of task_k] (out_k0) {\lstinline{o0}};
        \node [below=0.5cm of task_k] {\lstinline{...}};
        \node [circle,draw,below right=0.5cm and 0.5cm of task_k] (out_kn) {\lstinline{on}};
        
        \path [->,>=stealth] (k0) edge (task_k);
        \path [->,>=stealth] (kn) edge (task_k);
        \path [->,>=stealth] (task_k) edge (out_k0);
        \path [->,>=stealth] (task_k) edge (out_kn);
        
        \node [left=1cm of k0] (thread) {Thread $t$};
        \draw [snake=zigzag] (thread) -- (-3,-1);
        \path [->,>=latex,dashed] (task_k) edge[out=west,in=east] node [midway, below] (mid) {} (thread);
        
        \node [below left=0.25cm and -0.25cm of out_k0] (mapping) {\lstinline{(int)mapping(K k)}};
        \path [->,>=latex,dashed] (mapping) edge (mid);
        
        \draw [decorate,decoration={brace,amplitude=4}] (1.5,1.3) -- (1.5,0.6) node [] (brace0) {};
        \node [above right=-0.05cm and 0cm of brace0] {\lstinline{(int)indegree(K k)}};
        
        \draw [decorate,decoration={brace,amplitude=5}] (1.5,0.2) -- (1.5,-1.2) node [] (brace1) {};
        \node [above right=0.35cm and 0cm of brace1] {\lstinline{(void)run(K k)}};
        
    \end{tikzpicture}
    \caption{The \lstinline{Taskflow<K>} API. 
    \lstinline{(int)indegree(K k)} returns the number of incoming dependencies of task \lstinline{k}.
    \lstinline{(void)run(K k)} indicates what function to run.
    \lstinline{(int)mapping(K k)} returns what thread the task should be mapped \added{(but not bound)} to.
    }
    \label{fig:taskflow}
\end{figure}

\paragraph{Taskflow}
A \lstinline{Taskflow<K>} \lstinline{tf} (for some index space \lstinline{K}, typically an 
integer or a tuple of integers) represents a Parametrized Task Graph. It is created using \lstinline{Taskflow<K> tf(&tp)} where \lstinline{tp} is a \lstinline{Threadpool}.
It is responsible for managing task dependencies and
automatically inserting tasks in \lstinline{tp} when ready.
At least three functions have to be provided:
\begin{itemize}
    \item \lstinline{(int)indegree(K k)} returns the number of dependencies for task \lstinline{k}.
    \item \lstinline{(void)task(K k)} indicates what task \lstinline{k} should be doing when running.
            Typically this is some computational routine followed by the trigger of other tasks. 
            For instance task \lstinline{k1} can fulfill one dependency of task \lstinline{k2} by 
            \lstinline{tf.fulfill_promise(k2)}.
    \item \lstinline{(int)mapping(K k)} indicates what thread should task \lstinline{k} be 
            initially mapped to.
\end{itemize}
\added{In general, tasks can be stolen between threads to avoid starvation. This is done using a work stealing algorithm. \lstinline{tf.set_binding(binding)} can be used to make some tasks bound to their thread. Optional priorities can also be provided through \lstinline{tf.set_priority(priority)}. Finally, \lstinline{tf.fulfill_promise(k)} is used to fulfill one of the dependencies of task \lstinline{k} on Taskflow \lstinline{tf}. See \Cref{fig:taskflow}.}

\subsubsection{Distributed memory components} \label{sec:api_distributed}

Active Messages (AMs) are used to allow tasks on rank $a$ to trigger tasks on rank $b \neq a$ 
without rank $b$ explicitly waiting for messages.

An AM is a pair \lstinline{(function, payload)}.
When an AM is sent from rank $a$ to rank $b$, the payload is sent through the network, 
and upon arrival, the function (with the associated payload passed as argument) is run on 
the receiver rank. 
This allows for instance to store the payload at some location in local memory and then trigger tasks.

\paragraph{Active message}
An \lstinline{ActiveMsg<Ps...>} \lstinline{am} pairs a function \lstinline{(void)fun(Ps... ps)} and a payload \lstinline{ps}.
\added{Note that \lstinline{Ps...} is a variadic template: different types can be used as arguments. A \lstinline{view<T>} can be used to identify a memory buffer (i.e., a pointer and a length) and is built as
\lstinline{view<T> v(pointer, num_elements)}.}

The AM can be sent to rank \lstinline{dest} over the network using \lstinline{am->send(dest, ps...)}.
When sent, the payload is serialized on the sender, sent over the network, 
deserialized on the receiver and the function is run as \lstinline{fun(ps...)}.
The payloads are always serialized in a temporary buffer by the library. 
As such, the user-provided arguments can be immediately reused or modified 
as soon as \lstinline{send} returns.
\lstinline{am->send} is thread-safe and can be called by any thread.

\Ttor{} also provides \emph{large} active messages. 
A large AM can be used to avoid temporarily copying large buffers.
A large AM payload is made of one \lstinline{view<T>} and a series of arguments \lstinline{Ps...}.
The view will be sent and received directly without any extra copy.
\added{It is associated with three functions: 
(1) a function to be run on the receiver rank that returns a pointer to a user-allocated buffer, where the data will be stored;
(2) a function to be run on the receiver rank to process the data upon arrival;
(3) a function to be run on the sender rank when the buffer on the sender side can be reused.
This is an important feature to avoid costly copies and/or when memory use is constrained.}

\paragraph{Communicator}
A \lstinline{Communicator} \lstinline{comm} is a C++ factory to create AMs and is responsible 
for sending, receiving and running AMs. 
\lstinline{Communicator comm(mpi_comm)} creates a communicator using the \lstinline{mpi_comm} MPI communicator.
An AM can then be created by \lstinline[breaklines=true,breakatwhitespace=true]{am = comm.make_active_msg(f)} where \lstinline{f} is a \lstinline{(void)f(Ps...)} function.
AMs always have to be created in the same order on all ranks
because we need to create a consistent global indexing of all the AM that need to be run.


\subsubsection{Example}

The following shows how the different components can be used together. This assumes \lstinline{compute(k)} does the computation related to task \lstinline{k}. In addition, \lstinline{mapping(k)} returns a thread for task \lstinline{k} (which is typically \lstinline{k 

\begin{lstlisting}[basicstyle=\footnotesize\ttfamily]
/** Initialize structures **/
Communicator comm(MPI_COMM_WORLD);
Threadpool tp(n_threads, &comm);
Taskflow<int> tf(&tp);
/** Create active message **/
am = comm.make_active_msg(
  [&](int d, int k, payload pk) {
    data[k] = pk;
    tf.fulfill_promise(d);
  });
/** Define Taskflow **/
tf.set_mapping(mapping);
tf.set_indegree(n_deps);
tf.set_run([&](int k) {
  compute(k);
  for (auto d : deps(k)) {
    int dest = task_2_rank(d);
    if (dest == my_rank) {
      tf.fulfill_promise(d);
    } else {
      am->send(dest, d, k, data[k]);
    }
  }
});
/** Start initial tasks **/
for (auto k : initial_tasks)
  tf.fulfill_promise(k);
/** Wait for completion **/
tp.join();
\end{lstlisting}

%% file: implementation.tex
\subsection{Implementation Details}

\subsubsection{Taskflow and threadpool}

The threadpool is implemented with two
\lstinline{std::priority_queue<Task*>} per thread, storing the ready-to-run tasks. 
Since some tasks can be stolen and others not, each thread has two queues.
The priority queues are protected using \lstinline{std::mutex} so that tasks can be inserted into a thread queue by any other thread.

One of the main goals of the \lstinline{Taskflow<K>} implementation is to support arbitrary task flows with keys belonging to any domain. Hence, we store dependencies in a \lstinline{std::unordered_map<K,int>}. Furthermore, to avoid having one central map storing all dependencies (whose access needs to be serialized), the map is distributed across threads. Task's dependencies are split among the threads using the mapping function: the dependency count of task \lstinline{k} is stored in the map associated to thread \lstinline{mapping(k)}. Each distributed map is always accessed by the same thread, preventing data races.

\subsubsection{Active messages and communication thread}

Active messages (AM) are implemented by registering functions on every rank in the same order.
Each AM then has a unique ID shared across ranks. 
This ID is later used to retrieve the function on the receiver side.

Communication is performed using MPI non-blocking sends and receives.
The \lstinline{Communicator} maintains three queues:
\begin{enumerate}
    \item a queue of serialized and ready-to-send messages;
    \item a queue of send messages, to be later freed when the associated send completes;
    \item a queue of receive messages, to be later run and freed when the associated receive completes.
\end{enumerate}

On the sender side, when sending (thread-safe) an active message \lstinline{am->send(dest, ps...)}, the various arguments \lstinline{ps...} are first serialized into a buffer, along with the AM ID. The buffer is placed in a queue in the communicator. When calling \lstinline{progress()}, that buffer will eventually be sent using \lstinline{MPI_Isend} and later freed when the send has completed.

On the receiver side, calling \lstinline{progress()} performs the following:
\begin{enumerate}
    \item As long as it succeeds, it calls \lstinline{MPI_Iprobe} to probe for incoming messages and
            (1) retrieves the message size using \lstinline{MPI_Getcount},
            (2) allocates a buffer and
            (3) receives the message using \lstinline{MPI_Irecv}.
    \item It goes through all received messages and tests for completion with \lstinline{MPI_Test}.
          If it succeeds
            (1) it retrieves the AM using the ID from the buffer and
            (2) deserializes the buffer, passes the arguments to the user function and runs the user function.
\end{enumerate}
MPI tags are used to distinguish
(1) messages of size smaller or larger than $2^{31}$ bytes, and
(2) regular and large AMs.

%% file: completion.tex
\subsubsection{Distributed completion algorithm}
\label{sec:distributed_completion}

We now discuss the distributed algorithm to determine completion. We present the algorithm along with a proof of correctness. The difficulty in detecting completion lies in the fact that even if all taskflows are idle, the program may not be finished since active messages (AM) may still be in-flight. An example of a flawed strategy is to request that all ranks send an \lstinline{IDLE} signal to one rank when they have no tasks running. This strategy will lead to early termination of the program in many cases. Hence, detecting completion is non-trivial in a distributed setting.

\paragraph{Completion}

In the following, we will consider a series of events such as queuing and processing messages, checking certain conditions, etc. 
Within a thread we assume a total ordering between events which lets us associate each of them with a unique real number which we informally call ``time''.
We consider a program with two threads per rank: a main (MPI) thread responsible for MPI communication (asynchronous sends and receives) and AMs, and a worker thread responsible for executing all the user-defined tasks (in practice, the worker thread may be in fact a thread pool, but this is not relevant).

We say that an AM is \textbf{queued} on a sending rank when it is issued either by the worker or the main thread. When issued by a worker, we assume that queueing always finishes before the completion of the enclosing task. An AM is \textbf{processed} on the receiving rank by the main thread. We assume that if an AM results in a task being inserted in the task queue of the worker thread, this insertion must complete before the end of the enclosing AM.

To define our ordering between ranks, we assume that if a message is queued at time $t$ and processed at time $t'$ then $t' > t$. 
We assume that messages that are queued are eventually processed if the network and all ranks are idle except for handling these messages (progress guarantee), and that all communications are non blocking (no deadlocks are possible). \Ttor{} satisfies those assumptions by construction.

\begin{definition}[Completion]
We say that $\{t_a\}_a$ is completion time sequence if:
\begin{itemize}
    \item Rank $a$ is idle at time $t_a$ for all $a$;
    \item For any pair of ranks $(a,b)$ and all AMs from $a$ to $b$, all AMs queued before $t_a$ have been processed on $b$ before $t_b$.
\end{itemize}
\end{definition}
One can prove (omitted here) that this definition implies the intuitive definition of completion,
which is that, after $t_a$, if we keep the program running, rank $a$ remains idle forever.


\paragraph{Completion algorithm}

The algorithm is based on making sure, after all ranks are idle, that the number of messages sent is equal to the number of messages received. 
For this verification to work, we need to proceed in two steps, leading to the following definition:
\begin{definition}[Synchronization time]
    Assume that for all ranks $a$, we have defined a pair of times $(t^-_a,t^+_a)$ with $t^-_a < t^+_a$. We say that $\bar t$ is a synchronization time for $(t^-_a,t^+_a)$ if 
    $$ t^-_a < \bar t < t^+_a, \quad \text{for all $a$}$$
\end{definition}
Before giving the exact algorithm, we prove a sufficient condition to establish completion. 
\begin{lemma} \label{lemma:t_minus_t_plus}
    \added{Let $p_a(t)$ (resp.\ $q_a(t)$) be the number of processed (resp.\ queued) AMs on rank $a$ at time $t$. Assume that there exists a synchronization time $\bar t$ for $(t^-_a,t^+_a)$ and that for all $a$}
    \begin{itemize}
        \item the worker thread on rank $a$ is idle at $t_a^-$;
        \item $\bar p_a = p_a(t_a^-) = p_a(t_a^+)$ (no new processed AM between $t_a^-$ and $t_a^+$);
        \item $\bar q_a = q_a(t_a^-) = q_a(t_a^+)$ (no new queued AM between $t_a^-$ and $t_a^+$);
        \item $\sum_a \bar q_a = \sum_a \bar p_a$.
    \end{itemize}
    Then the sequence $\{t_a^-\}_a$ is a completion time sequence for the execution.
\end{lemma}

\begin{IEEEproof}
    Let us first prove that rank $a$ is idle during the entire period $[t_a^-,t_a^+]$. Rank $a$ is idle at $t_a^-$. Since $p_a(t_a^-) = p_a(t_a^+)$, no AM was processed at any time $t \in [t_a^-,t_a^+]$. So no tasks may have been inserted in the worker task queue by the main thread. Hence, rank $a$ is idle during $[t_a^-,t_a^+]$.

    Second, because $p_a(t_a^-) = p_a(t_a^+)$ and $q_a(t_a^-) = q_a(t_a^+)$, we necessarily have that
    $$ p_a(\bar t) = p_a(t_a^-) = p_a(t_a^+), \quad q_a(\bar t) = q_a(t_a^-) = q_a(t_a^+). $$
    This is because $p_a$ and $q_a$ are increasing functions of time and $t_a^- < \bar t < t_a^+$. Therefore: $\sum_a \bar q_a(\bar t) = \sum_a \bar p_a(\bar t)$. The key is that this is true at the synchronization time $\bar t$.

    Consider now a message $m$ that is contributing to $\sum_a \bar q_a(\bar t)$ and $\sum_a \bar p_a(\bar t)$. It is not possible that $m$ contributes $+1$ to $\sum_a \bar p_a(\bar t)$ (e.g., it has been counted as processed) while contributing 0 to $\sum_a \bar q_a(\bar t)$ (e.g., it has not been counted as queued). This is because the process time is always strictly greater than the queuing time and we are evaluating the terms at the synchronization time $\bar t$.

    From this, assume now that $m$ contributes $+1$ to $\sum_a \bar q_a(\bar t)$ (queued) and 0 to $\sum_a \bar p_a(\bar t)$ (but not processed yet). Then we must have:
    $$ \sum_a \bar q_a(\bar t) > \sum_a \bar p_a(\bar t) $$
    This is because not other message $m'$ can ``restore'' the equality. This inequality is a contradiction.

    Therefore all messages queued have been processed. With the results above, $\{t_a^-\}$ is a completion time sequence.
\end{IEEEproof}

We now describe the algorithm.
Rank 0 will be responsible to detect completion by synchronizing ($\bar t$) with other ranks $r > 0$.
\textbf{When a rank is idle}, the main thread on all ranks does the following.
\begin{enumerate}
    \item All ranks $r$ continuously monitor $q_r(t)$ and $p_r(t)$ (which only contain the user's AM count and not the messages used in the completion algorithm). If at a time $t^-_r$ those values differ from the latest observed ones, rank $r$ sends a message \lstinline{COUNT} $= (r, q_r(t^-_r), p_r(t^-_r))$ to rank $0$ with those updated counts.
    \item Rank $0$ continuously observes the latest received counts. Since $q_r(\cdot)$ and $p_r(\cdot)$ are non-decreasing it is enough to consider the greatest received counts and discard the others. If at time $\tilde t$ (implemented as an always increasing integer counter), $\sum_r q_r(t^-_r) = \sum_r p_r(t^-_r)$ and that sum is different from the latest observed sum, rank $0$ sends a \lstinline{REQUEST} $= (q_r(t^-_r),p_r(t^-_r), \tilde t)$ message back to all ranks $r > 0$.
    \item All ranks $r$ continuously monitor the \lstinline{REQUEST} messages from rank $0$. They process the one with the largest $\tilde t$, and discard the others. At time $t^+_r$, if $q_r(t^-_r) = q_r(t^+_r)$ and $p_r(t^-_r) = p_r(t^+_r)$, they send a \lstinline{CONFIRMATION} = $(\tilde t)$ back to rank $0$.
    \item Rank $0$ continuously observes the received \lstinline{CONFIRMATION}. If all ranks replied with the latest $\tilde t$, the program has completed. Rank $0$ then sends a \lstinline{SHUTDOWN} message to all ranks.
    \item All ranks $r$ continuously listen to the \lstinline{SHUTDOWN} message. When received, the program has completed and rank $r$ terminates.
\end{enumerate}

Note that although we write the algorithm as a sequence from 1 to 5, the word ``continuously'' indicates that this is implemented as a loop which keeps attempting to perform each step until \lstinline{SHUTDOWN} is received.



We proved the following two theorems (proof is omitted but is based on the results described above, with the assumptions provided at the beginning).
\begin{theorem}[Correctness]
    The \lstinline{SHUTDOWN} message is sent if and only if completion has been reached.
\end{theorem}
    

The second property guarantees that \lstinline{CONFIRMATION} is sent in finite time. For example, if the number of message is potentially unbounded, messages from some ranks could always be prioritized, preventing any progress from other ranks, and the algorithm may never terminate.

\begin{theorem}[Finiteness]
    The completion protocol is guaranteed to send \lstinline{CONFIRMATION} in finite time.
\end{theorem}



%% file: benchmarks.tex
\section{Benchmarks}
\label{sec:benchmarks}

In this section, we present benchmarks comparing \Ttor{} to OpenMP, StarPU, and ScaLAPACK.

We start with micro-benchmarks to validate the low overhead of the shared memory component. This is only used to verify that the task-based management overhead is comparable, and sometimes better, to other runtime systems.

\added{We then apply \Ttor{} (with its distributed component) to two classical linear algebra problems. In those sections, the goal is to compare a sequential enumeration of the DAG (STF) as implemented in StarPU versus the PTG approach as implemented in \Ttor{}. We note in particular that it is possible to modify the StarPU code such that the DAG is parallelized in a manner close to \Ttor{}. Similarly several optimizations in \Ttor{} are possible but were not explored for this paper (memory management, task insertion, communication). Therefore, these benchmarks cannot be interpreted as measuring the peak performance of either runtime.}


\added{In all cases, experiments are run on a cluster equipped with dual-sockets and 16 cores Intel(R) Xeon(R) CPU E5-2670 0 @ 2.60GHz with 32GB of RAM per node. Intel Compiler (\lstinline{icpc (ICC) 19.1.0.166 20191121}) and Intel MPI are used with Intel MKL (version \lstinline{2020.0.166}) for BLAS, LAPACK and ScaLAPACK. We use StarPU version \lstinline{1.3.2}. We assign one MPI rank per node. \Ttor{}'s code, including benchmarks, is available at {\small \lstinline[breaklines=true]{github.com/leopoldcambier/tasktorrent}}. StarPU and ScaLAPACK's benchmarks are available at {\small \lstinline[breaklines=true]{github.com/leopoldcambier/tasktorrent_paper_benchmarks}}.}

%% file: micro_benchmarks.tex
\subsection{Micro-benchmarks}

We first perform a series of micro benchmarks to validate the low overhead of the shared memory component of the runtime. In the following, we average timings across 25 runs. In every case, the standard deviation was recorded as well, to estimate the variability of the measurement. In most cases, it was negligible and we don't report it. In all cases, we pick a number of tasks so that the total runtime is about 1 second.

\subsubsection{No-dependencies overhead}

\begin{figure}
  \centering
  \subfloat[][\label{sfig:benchmark_no_deps_a}
  TTOR's overhead (no dependencies) measurement. 
  Task insertion time is not included. Numbers indicate \lstinline{spin_time}.]{
  \begin{tikzpicture}
    \begin{axis}[
      font=\footnotesize, ytick={0,0.5,1.0},
      xlabel={Num.\ threads}, ylabel={Efficiency},
      width=4.5cm, ylabel near ticks, ymax=1.2, ymin=-0.1,
      ]
      \addplot+[mark=*,mark options={mark size=1pt,solid},red]    table[x=n_threads,y=efficiency_mean,y error=efficiency_std,discard if not={sleep_time}{1.000000e-04}]{micro_benchmarks/no_deps.dat};
      \addplot+[mark=*,mark options={mark size=1pt,solid},blue]   table[x=n_threads,y=efficiency_mean,y error=efficiency_std,discard if not={sleep_time}{1.000000e-05}]{micro_benchmarks/no_deps.dat};
      \addplot+[mark=*,mark options={mark size=1pt,solid},brown]  table[x=n_threads,y=efficiency_mean,y error=efficiency_std,discard if not={sleep_time}{1.000000e-06}]{micro_benchmarks/no_deps.dat};
      \node (text) at (axis cs:12.5,0.3) {\textcolor{brown}{1 $\mu$s}};
      \node (text) at (axis cs:12.5,0.75) {\textcolor{blue}{10 $\mu$s}};
      \node (text) at (axis cs:12.5,1.1) {\textcolor{red}{100 $\mu$s}};
    \end{axis}
  \end{tikzpicture}
  }\;
  \subfloat[][\label{sfig:benchmark_no_deps_b}
  Overhead (no dependencies) comparisons. 
  Task insertion time is included.
  Solid is \lstinline{spin_time} = 100 $\mu$s; dashed is 10 $\mu$s;
  * is StarPU with direct task insertion (Task) and STF semantics (STF).]{
  \begin{tikzpicture}
    \begin{axis}[
      font=\footnotesize,
      xlabel={Num.\ threads}, ylabel={Efficiency},
      width=4.5cm, ylabel near ticks, ymax=1.2, ymin=-0.1,
    ]
    \addplot[mark=*,mark options={mark size=1pt,solid},red,dashed,forget plot]    table[x=n_threads,y=efficiency_mean,y error=efficiency_std,discard if not={sleep_time}{1.000000e-05},discard if not={test}{ttor_wait_time_insertion1}]{micro_benchmarks/no_deps_comp.dat};
    \addplot[mark=*,mark options={mark size=1pt,solid},red,]                      table[x=n_threads,y=efficiency_mean,y error=efficiency_std,discard if not={sleep_time}{1.000000e-04},discard if not={test}{ttor_wait_time_insertion1}]{micro_benchmarks/no_deps_comp.dat};
    \addplot[mark=*,mark options={mark size=1pt,solid},blue,dashed,forget plot]   table[x=n_threads,y=efficiency_mean,y error=efficiency_std,discard if not={sleep_time}{1.000000e-05},discard if not={test}{omp_wait}]{micro_benchmarks/no_deps_comp.dat};
    \addplot[mark=*,mark options={mark size=1pt,solid},blue]                      table[x=n_threads,y=efficiency_mean,y error=efficiency_std,discard if not={sleep_time}{1.000000e-04},discard if not={test}{omp_wait}]{micro_benchmarks/no_deps_comp.dat};
    \addplot[mark=*,mark options={mark size=1pt,solid},black,dashed,forget plot]  table[x=n_threads,y=efficiency_mean,y error=efficiency_std,discard if not={sleep_time}{1.000000e-05},discard if not={test}{startpu_wait}]{micro_benchmarks/no_deps_comp.dat};
    \addplot[mark=*,mark options={mark size=1pt,solid},black]                     table[x=n_threads,y=efficiency_mean,y error=efficiency_std,discard if not={sleep_time}{1.000000e-04},discard if not={test}{startpu_wait}]{micro_benchmarks/no_deps_comp.dat};
    \addplot[mark=*,mark options={mark size=1pt,solid},brown,dashed,forget plot]  table[x=n_threads,y=efficiency_mean,y error=efficiency_std,discard if not={sleep_time}{1.000000e-05},discard if not={test}{startpu_wait_stf}]{micro_benchmarks/no_deps_comp.dat};
    \addplot[mark=*,mark options={mark size=1pt,solid},brown]                     table[x=n_threads,y=efficiency_mean,y error=efficiency_std,discard if not={sleep_time}{1.000000e-04},discard if not={test}{startpu_wait_stf}]{micro_benchmarks/no_deps_comp.dat};
    \node (text) at (axis cs:9,0.1) {\textcolor{brown}{* (STF)}};
    \node [fill=white,inner sep=1pt] (text) at (axis cs:8.3,0.68) {\textcolor{black}{* (Task)}};
    \node (text) at (axis cs:15,0.58) {\textcolor{blue}{OMP}};
    \node (text) at (axis cs:15,0.85) {\textcolor{red}{\Ttor{}}};
    \end{axis}
  \end{tikzpicture}
  }

  \caption{Shared memory serial overhead, as a function of the number of threads \lstinline{nthreads} (x-axis) and the task time \lstinline{spin_time} (various lines). 
  The plots show the mean across 25 runs.}
  \label{fig:benchmark_no_deps}
\end{figure}
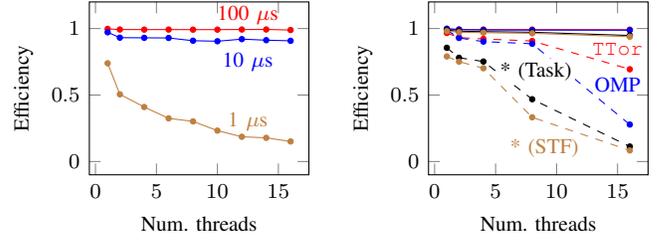

We begin with an estimation of the ``serial'' overhead of \Ttor{}'s shared memory runtime. 
We start \lstinline{ntasks} tasks, without any dependencies, and assign them in a round-robin fashion to the \lstinline{nthreads} threads.
Each task is only spinning for \lstinline{spin_time} seconds. 
As such, the total ideal time is \lstinline{spin_time} $\times$ \lstinline{ntasks} $/$ \lstinline{nthreads}.
\Cref{fig:benchmark_no_deps} shows the efficiency as a function of \lstinline{nthreads} and \lstinline{spin_time}.
Given a total wall clock time of \lstinline{run_time}, efficiency is defined as \lstinline{run_time} $\times$ \lstinline{nthreads} $/$ (\lstinline{spin_time} $\times$ \lstinline{ntasks}).
\lstinline{ntasks} is chosen so that \lstinline{run_time} is around 2 seconds.

\Cref{sfig:benchmark_no_deps_a} shows results for \Ttor{}'s only, 
where we do \textbf{not} measure task insertion, i.e., we evaluate
\begin{lstlisting}[basicstyle=\footnotesize\ttfamily]
for(int k = 0; k < n_tasks; k++) {
  tf.fulfill_promise(k);
}
tp.start(); // Start measuring time
tp.join();  // Stop measuring time
\end{lstlisting}
We see that the runtime has negligible impact for tasks $\approx$ 100$\mu$s, 
and it becomes significant around 1 $\mu$s where overhead dominates.

We then compare it to OpenMP and StarPU in \Cref{sfig:benchmark_no_deps_b} where, to make the comparison fair, insertion time \textbf{is} measured 
(which reduces the maximum possible efficiency, as the insertion is sequential).
\begin{lstlisting}[basicstyle=\footnotesize\ttfamily]
tp.start(); // Start measuring time
for(int k = 0; k < n_tasks; k++) {
  tf.fulfill_promise(k);
}
tp.join();  // Stop measuring time
\end{lstlisting}
We note that this is a spurious consequence of creating tasks with no dependencies. In practice the insertion is done by other tasks, themselves executing in parallel. We evaluate StarPU both using ``direct'' task insertion (``Task''), as well as using the STF approach (``STF''). In the STF approach, each independent task is associated with an artificial independent read-write piece of data. We see that for very small tasks $<$ 10$\mu s$, overhead is significant but comparable for all runtimes.

\subsubsection{Many dependencies overhead}

We then estimate the overhead when dependencies are involved.
Consider a 2D array of \lstinline{nrows} $\times$ \lstinline{ncols} tasks, with \lstinline{ndeps} dependencies between task $(i,j)$ and $((i+k)\text{\%\lstinline{nrows}},j+1)$ for $0 \leq k <$ \lstinline{ndeps}.
Again, tasks are spinning for \lstinline{spin_time} seconds and, in \Ttor{}, task $(i,j)$ is assigned to thread $i \%$ \lstinline{nthreads}.

\added{Since this is not easily implementable in OpenMP,} we only compare \Ttor{} with StarPU. In the ``Task'' version, tasks are directly inserted, and their dependencies are explicitly expressed. In the STF approach, we register data for every $(i,j)$ task and that data is used to create dependencies with the tasks in the next column. We note that StarPU STF has the constraint that the number of input data buffers for a given task should normally be known at compile time, which makes it not well-suited for this benchmark.

\begin{figure}
  \centering
  \begin{tikzpicture}
    \begin{groupplot}[
      group style={
          group size=2 by 3,
      }
    ]
    \nextgroupplot[
      title={\lstinline{spin_time} = 100$\mu$s},
      ylabel={TTOR},
      ylabel style={align=center,at={(-0.2,0.5)}},
      width=4cm, font=\footnotesize, ymin=0.3, ymax=1.1,
    ]
    \addplot+[mark=*,mark options={mark size=1pt,solid},blue  ]   table[x=n_threads,y=efficiency_mean,y error=efficiency_std,discard if not={n_edges}{1},discard if not={sleep_time}{1.000000e-04},discard if not={test}{ttor_deps}]{micro_benchmarks/deps_comp.dat};
    \addplot+[mark=*,mark options={mark size=1pt,solid},red   ]  table[x=n_threads,y=efficiency_mean,y error=efficiency_std,discard if not={n_edges}{2},discard if not={sleep_time}{1.000000e-04},discard if not={test}{ttor_deps}]{micro_benchmarks/deps_comp.dat};
    \addplot+[mark=*,mark options={mark size=1pt,solid},purple]  table[x=n_threads,y=efficiency_mean,y error=efficiency_std,discard if not={n_edges}{4},discard if not={sleep_time}{1.000000e-04},discard if not={test}{ttor_deps}]{micro_benchmarks/deps_comp.dat};
    \addplot+[mark=*,mark options={mark size=1pt,solid},black ]  table[x=n_threads,y=efficiency_mean,y error=efficiency_std,discard if not={n_edges}{8},discard if not={sleep_time}{1.000000e-04},discard if not={test}{ttor_deps}]{micro_benchmarks/deps_comp.dat};
    \addplot+[mark=*,mark options={mark size=1pt,solid},brown ]  table[x=n_threads,y=efficiency_mean,y error=efficiency_std,discard if not={n_edges}{16},discard if not={sleep_time}{1.000000e-04},discard if not={test}{ttor_deps}]{micro_benchmarks/deps_comp.dat};
    \nextgroupplot[
      title={\lstinline{spin_time} = 1ms},
      legend entries={1 dep, 2 deps, 4 deps, 8 deps, 16 deps, 32 deps},
      legend pos=outer north east, 
      ytick={},
      width=4cm, font=\footnotesize, ymin=0.3, ymax=1.1,
    ]
    \addplot+[mark=*,mark options={mark size=1pt,solid},blue  ] table[x=n_threads,y=efficiency_mean,y error=efficiency_std,discard if not={n_edges}{1},discard if not={sleep_time}{1.000000e-03},discard if not={test}{ttor_deps}]{micro_benchmarks/deps_comp.dat};
    \addplot+[mark=*,mark options={mark size=1pt,solid},red   ] table[x=n_threads,y=efficiency_mean,y error=efficiency_std,discard if not={n_edges}{2},discard if not={sleep_time}{1.000000e-03},discard if not={test}{ttor_deps}]{micro_benchmarks/deps_comp.dat};
    \addplot+[mark=*,mark options={mark size=1pt,solid},purple] table[x=n_threads,y=efficiency_mean,y error=efficiency_std,discard if not={n_edges}{4},discard if not={sleep_time}{1.000000e-03},discard if not={test}{ttor_deps}]{micro_benchmarks/deps_comp.dat};
    \addplot+[mark=*,mark options={mark size=1pt,solid},black ] table[x=n_threads,y=efficiency_mean,y error=efficiency_std,discard if not={n_edges}{8},discard if not={sleep_time}{1.000000e-03},discard if not={test}{ttor_deps}]{micro_benchmarks/deps_comp.dat};
    \addplot+[mark=*,mark options={mark size=1pt,solid},brown ] table[x=n_threads,y=efficiency_mean,y error=efficiency_std,discard if not={n_edges}{16},discard if not={sleep_time}{1.000000e-03},discard if not={test}{ttor_deps}]{micro_benchmarks/deps_comp.dat};
    \nextgroupplot[
      yshift=0.5cm,
      ylabel={*PU Task},
      ylabel style={align=center,at={(-0.2,0.5)}},
      width=4cm, font=\footnotesize, ymin=0.3, ymax=1.1,
    ]
    \addplot+[mark=*,mark options={mark size=1pt,solid},blue  ] table[x=n_threads,y=efficiency_mean,y error=efficiency_std,discard if not={n_edges}{1},discard if not={sleep_time}{1.000000e-04},discard if not={test}{starpu_deps}]{micro_benchmarks/deps_comp.dat};
    \addplot+[mark=*,mark options={mark size=1pt,solid},red   ] table[x=n_threads,y=efficiency_mean,y error=efficiency_std,discard if not={n_edges}{2},discard if not={sleep_time}{1.000000e-04},discard if not={test}{starpu_deps}]{micro_benchmarks/deps_comp.dat};
    \addplot+[mark=*,mark options={mark size=1pt,solid},purple] table[x=n_threads,y=efficiency_mean,y error=efficiency_std,discard if not={n_edges}{4},discard if not={sleep_time}{1.000000e-04},discard if not={test}{starpu_deps}]{micro_benchmarks/deps_comp.dat};
    \addplot+[mark=*,mark options={mark size=1pt,solid},black ] table[x=n_threads,y=efficiency_mean,y error=efficiency_std,discard if not={n_edges}{8},discard if not={sleep_time}{1.000000e-04},discard if not={test}{starpu_deps}]{micro_benchmarks/deps_comp.dat};
    \addplot+[mark=*,mark options={mark size=1pt,solid},brown ] table[x=n_threads,y=efficiency_mean,y error=efficiency_std,discard if not={n_edges}{16},discard if not={sleep_time}{1.000000e-04},discard if not={test}{starpu_deps}]{micro_benchmarks/deps_comp.dat};
    \nextgroupplot[
      yshift=0.5cm,
      ytick={},
      width=4cm, font=\footnotesize, ymin=0.3, ymax=1.1,
    ]
    \addplot+[mark=*,mark options={mark size=1pt,solid},blue  ] table[x=n_threads,y=efficiency_mean,y error=efficiency_std,discard if not={n_edges}{1},discard if not={sleep_time}{1.000000e-03},discard if not={test}{starpu_deps}]{micro_benchmarks/deps_comp.dat};
    \addplot+[mark=*,mark options={mark size=1pt,solid},red   ] table[x=n_threads,y=efficiency_mean,y error=efficiency_std,discard if not={n_edges}{2},discard if not={sleep_time}{1.000000e-03},discard if not={test}{starpu_deps}]{micro_benchmarks/deps_comp.dat};
    \addplot+[mark=*,mark options={mark size=1pt,solid},purple] table[x=n_threads,y=efficiency_mean,y error=efficiency_std,discard if not={n_edges}{4},discard if not={sleep_time}{1.000000e-03},discard if not={test}{starpu_deps}]{micro_benchmarks/deps_comp.dat};
    \addplot+[mark=*,mark options={mark size=1pt,solid},black ] table[x=n_threads,y=efficiency_mean,y error=efficiency_std,discard if not={n_edges}{8},discard if not={sleep_time}{1.000000e-03},discard if not={test}{starpu_deps}]{micro_benchmarks/deps_comp.dat};
    \addplot+[mark=*,mark options={mark size=1pt,solid},brown ] table[x=n_threads,y=efficiency_mean,y error=efficiency_std,discard if not={n_edges}{16},discard if not={sleep_time}{1.000000e-03},discard if not={test}{starpu_deps}]{micro_benchmarks/deps_comp.dat};
    \nextgroupplot[
      yshift=0.5cm,
      xlabel={Num.\ threads}, 
      ylabel={*PU STF},
      ylabel style={align=center,at={(-0.2,0.5)}},
      xlabel near ticks,
      width=4cm, font=\footnotesize, ymin=0.3, ymax=1.1, ylabel style={align=center},
    ]
    \addplot+[mark=*,mark options={mark size=1pt,solid},blue  ] table[x=n_threads,y=efficiency_mean,y error=efficiency_std,discard if not={n_edges}{1},discard if not={sleep_time}{1.000000e-04},discard if not={test}{starpu_deps_stf}]{micro_benchmarks/deps_comp.dat};
    \addplot+[mark=*,mark options={mark size=1pt,solid},red   ] table[x=n_threads,y=efficiency_mean,y error=efficiency_std,discard if not={n_edges}{2},discard if not={sleep_time}{1.000000e-04},discard if not={test}{starpu_deps_stf}]{micro_benchmarks/deps_comp.dat};
    \addplot+[mark=*,mark options={mark size=1pt,solid},purple] table[x=n_threads,y=efficiency_mean,y error=efficiency_std,discard if not={n_edges}{4},discard if not={sleep_time}{1.000000e-04},discard if not={test}{starpu_deps_stf}]{micro_benchmarks/deps_comp.dat};
    \addplot+[mark=*,mark options={mark size=1pt,solid},black ] table[x=n_threads,y=efficiency_mean,y error=efficiency_std,discard if not={n_edges}{8},discard if not={sleep_time}{1.000000e-04},discard if not={test}{starpu_deps_stf}]{micro_benchmarks/deps_comp.dat};
    \addplot+[mark=*,mark options={mark size=1pt,solid},brown ] table[x=n_threads,y=efficiency_mean,y error=efficiency_std,discard if not={n_edges}{16},discard if not={sleep_time}{1.000000e-04},discard if not={test}{starpu_deps_stf}]{micro_benchmarks/deps_comp.dat};
    \nextgroupplot[
      yshift=0.5cm,
      xlabel={Num.\ threads}, xlabel near ticks,
      ytick={},
      width=4cm, font=\footnotesize, ymin=0.3, ymax=1.1,
    ]
    \addplot+[mark=*,mark options={mark size=1pt,solid},blue  ] table[x=n_threads,y=efficiency_mean,y error=efficiency_std,discard if not={n_edges}{1},discard if not={sleep_time}{1.000000e-03},discard if not={test}{starpu_deps_stf}]{micro_benchmarks/deps_comp.dat};
    \addplot+[mark=*,mark options={mark size=1pt,solid},red   ] table[x=n_threads,y=efficiency_mean,y error=efficiency_std,discard if not={n_edges}{2},discard if not={sleep_time}{1.000000e-03},discard if not={test}{starpu_deps_stf}]{micro_benchmarks/deps_comp.dat};
    \addplot+[mark=*,mark options={mark size=1pt,solid},purple] table[x=n_threads,y=efficiency_mean,y error=efficiency_std,discard if not={n_edges}{4},discard if not={sleep_time}{1.000000e-03},discard if not={test}{starpu_deps_stf}]{micro_benchmarks/deps_comp.dat};
    \addplot+[mark=*,mark options={mark size=1pt,solid},black ] table[x=n_threads,y=efficiency_mean,y error=efficiency_std,discard if not={n_edges}{8},discard if not={sleep_time}{1.000000e-03},discard if not={test}{starpu_deps_stf}]{micro_benchmarks/deps_comp.dat};
    \addplot+[mark=*,mark options={mark size=1pt,solid},brown ] table[x=n_threads,y=efficiency_mean,y error=efficiency_std,discard if not={n_edges}{16},discard if not={sleep_time}{1.000000e-03},discard if not={test}{starpu_deps_stf}]{micro_benchmarks/deps_comp.dat};
    \end{groupplot}
  \end{tikzpicture}
  \caption{Efficiency vs.\ number of threads. Shared memory runtime dependency management overhead. The plots show the mean across 25 runs.}
  \label{fig:benchmark_with_deps}
\end{figure}
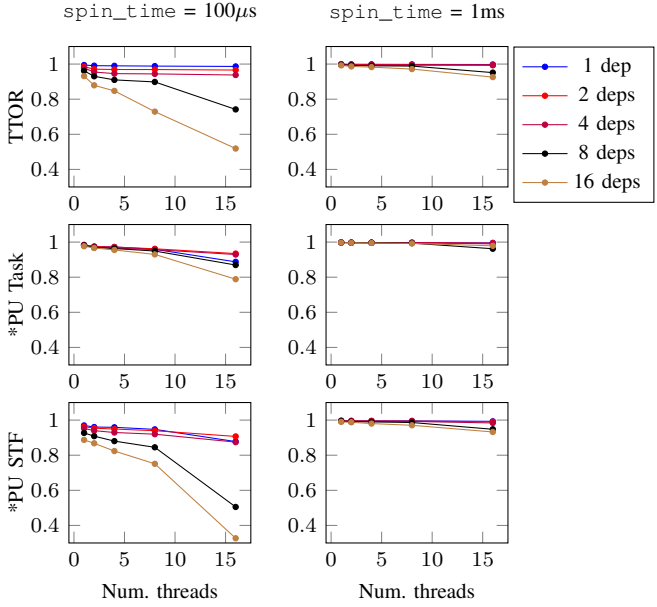

\Cref{fig:benchmark_with_deps} shows the results with \lstinline{nrows} set to 32.
We see that \Ttor{} is between StarPU ``Task'' and StarPU ``STF'', with similar overhead.
This validates the implementation.

\added{The conclusion of this section is that the overhead of \Ttor{} is comparable (and sometimes better) to OpenMP and StarPU.}

%% file: gemm.tex
\subsection{Distributed Matrix-matrix Product}

We now consider a distributed matrix-matrix multiplication problem (GEMM), i.e., given $A, B \in \mathbb{R}^{N \times N}$ compute $C = AB$.
We compare: 
\begin{itemize}
\item \Ttor{} with an algorithm using a 2D block cyclic mapping of blocks of size 256 to ranks, 
using the default (``small'') and large AMs;
\item \Ttor{} with an algorithm using a 3D mapping of blocks to ranks, tiled 
(every GEMM is single threaded, with a block size of 256) 
or not (every GEMM is a single large multithreaded BLAS). 
We use the DNS algorithm (see for instance \cite{grama2003introduction}) to map blocks to ranks. 
\item StarPU (with STF semantics, i.e., all ranks explore the full DAG) using a 2D block cyclic mapping of blocks of size 256 to ranks. Various scheduling strategies have been tried, without significant variation in runtime; the default local work stealing \lstinline{lws} is then used.
\item \added{ScaLAPACK using a 2D block cyclic mapping (with a block size of 256) with multithreaded BLAS. We note that ScaLAPACK is not a runtime and is not actively managing a task graph.}
\end{itemize}

The following code snippet shows the GEMM portion when using the 2D block cyclic data distribution. In this case, contributions $A_{ik} B_{kj}$ are ordered as function of $k$, i.e., $A_{ik} B_{kj}$ happens before $A_{i(k+1)} B_{(k+1)j}$. Furthermore, because of the 2D data distribution, the products $A_{ik} B_{kj}$ are mapped to a rank function of $(i,j)$ only and, as such, always happen on a same node. The mapping of tasks to thread may be any deterministic function of \lstinline{ikj}. In practice something as simple as \lstinline{ikj[0] 


\Cref{fig:gemm_scaling} presents strong and weak scalings results.
Scalings are done multiplying the number of rows and columns by 2 and/or the number of nodes by 8,
and the largest test case are matrices of size $32\,768$. We make multiple observations:
\begin{itemize}
  \item \Ttor{} benefits from the large messages (\Cref{sfig:gemm_2d}) over small ones, decreasing the total time by up to 30\%.
  \item \Ttor{} with large messages and StarPU using the 2D mapping have similar performance (\Cref{sfig:gemm_2d} vs \Cref{sfig:gemm_starpu}). \Ttor{} performs better than StarPU with small blocks (\Cref{sfig:gemm_blocksize}).
  \item \Ttor{} with the 3D mapping and the tiled algorithm has better performance than without (see \Cref{sfig:gemm_3d} as well as \Cref{sfig:gemm_3d_tiled} vs \Cref{sfig:gemm_3d_nontiled} for results on 8 nodes). This shows the importance of having a small task granularity, to increase overlap between communication and computation. It has however similar performance to the 2D mapping.
  \item Runtime-based implementations outperform ScaLAPACK (\Cref{sfig:gemm_scalapack}), showing the benefits of a task-based runtime system.
\end{itemize}
\Cref{sfig:gemm_blocksize} shows the impact of the block size on the runtime. 
We see that \Ttor{} is about 2.5x faster than StarPU at small sizes.
This highlights the advantages of a distributed DAG exploration. \added{We note that in this case small blocks are not optimal. However, GEMM is in some sense an ``easy'' benchmark since it offers a large amount of concurrency. Therefore, to stress the runtimes and observe measurable differences we need to deviate from the optimal GEMM settings. Although we could not investigate other algorithms for this paper, more complex applications would probably reveal additional differences between \Ttor{} and StarPU.\@}


\added{Finally, \Cref{sfig:gemm_concurrency} shows the efficiency of \Ttor{} (2D GEMM) as a function of the concurrency. Since the GEMMs are sequential as a function of $k$, \lstinline{num_blocks^2/n_cores} indicates how much parallelism is available per core. This represents the number of blocks that are processed on each core between communication steps. We see that efficiency decreases sharply at around 16 blocks per core.}

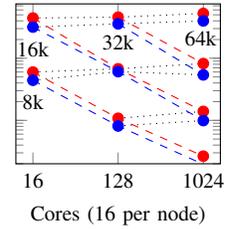
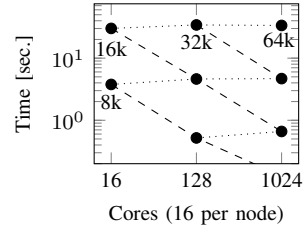
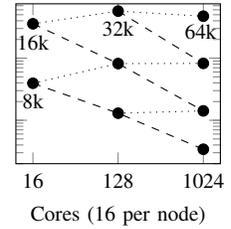
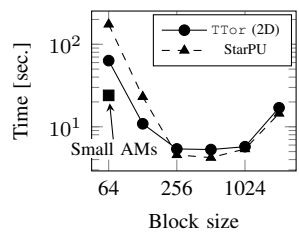
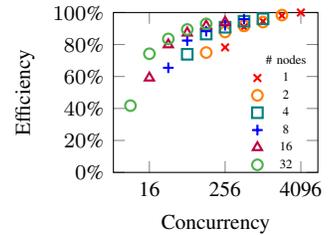
\begin{figure}
  \centering
  \subfloat[\label{sfig:gemm_3d_tiled}Tiled 3D DNS GEMM trace for 8 nodes. Every red line is 1 \Ttor{} thread using single-threaded BLAS.]{
    \begin{tikzpicture}
      \begin{axis}[
          font=\footnotesize,xmin=0,xmax=5,ymin=0,ymax=1,width=0.24\textwidth,
          ytick=\empty,yticklabels=\empty,xlabel={Time [sec.]},
        ]
        \addplot[] graphics[xmin=0,xmax=5,ymin=0,ymax=1] {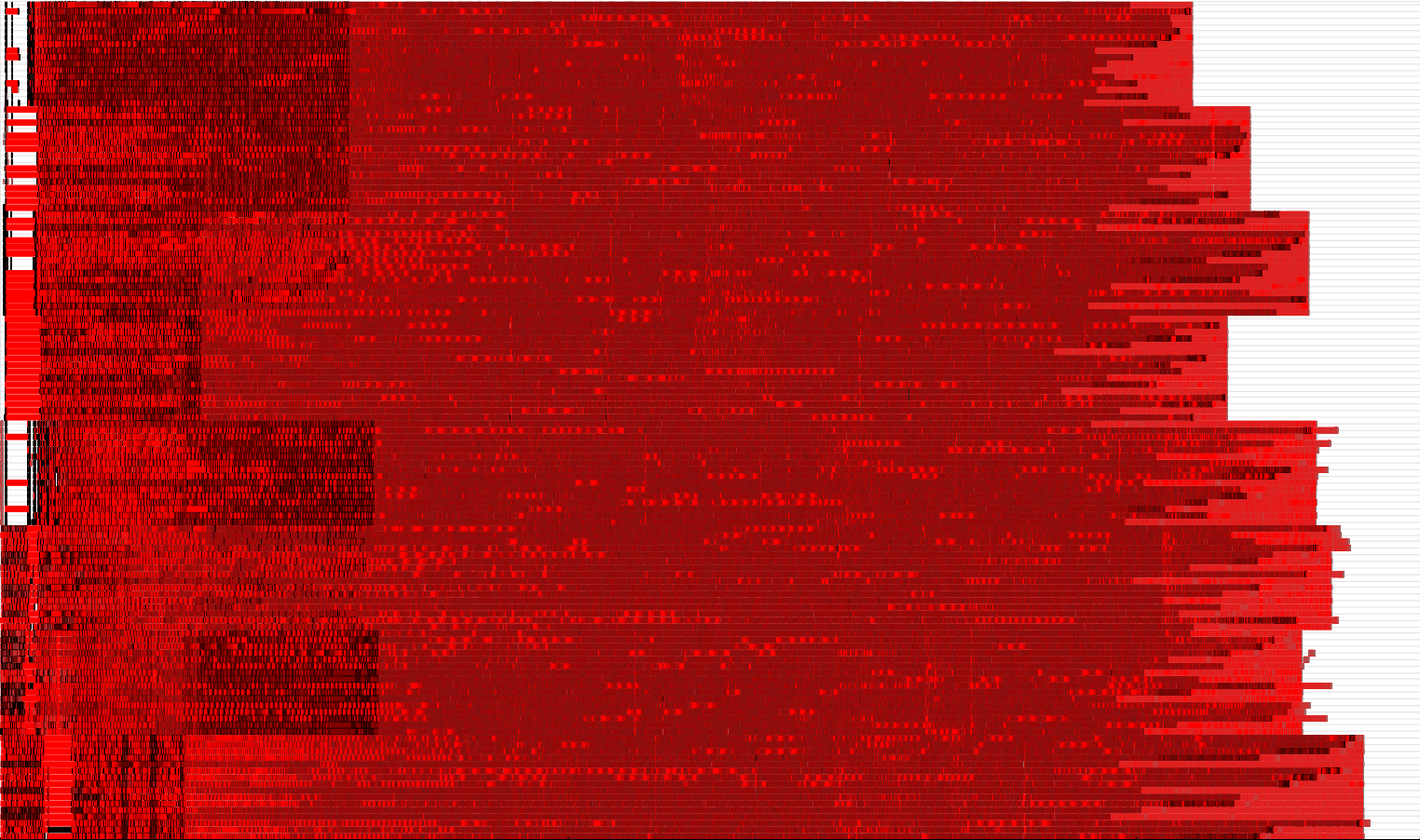};
      \end{axis}
    \end{tikzpicture}
  }\quad\quad
  \subfloat[\label{sfig:gemm_3d_nontiled}Non-tiled 3D DNS GEMM trace for 8 nodes. Every red line is 1 \Ttor{} thread using multithreaded BLAS.]{
    \begin{tikzpicture}
      \begin{axis}[
          font=\footnotesize,xmin=0,xmax=7,ymin=0,ymax=1,width=0.24\textwidth,
          ytick=\empty,yticklabels=\empty,xlabel={Time [sec.]}
        ]
        \addplot[] graphics[xmin=0,xmax=7,ymin=0,ymax=1] {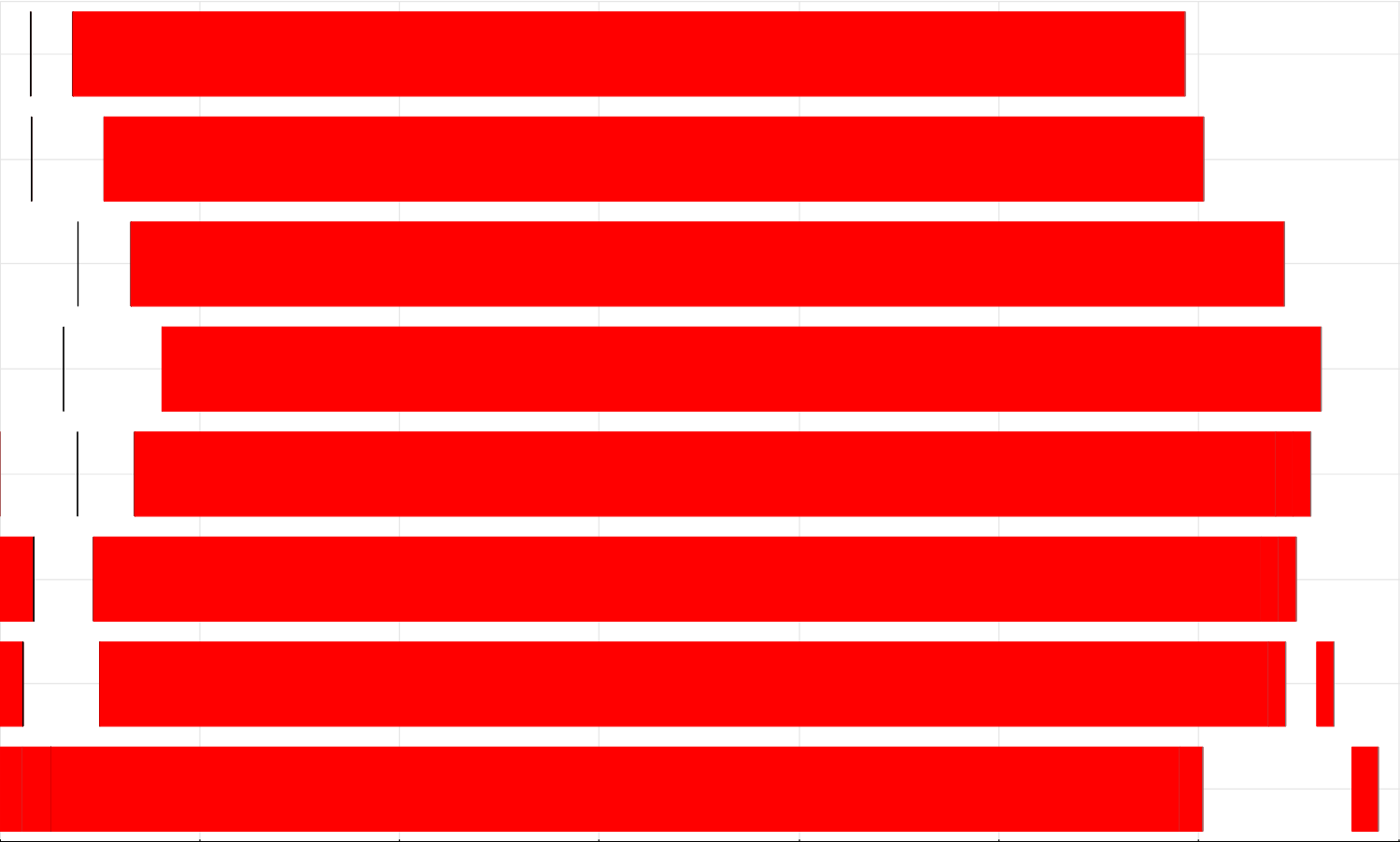};
      \end{axis}
    \end{tikzpicture}
  }\\
  \subfloat[\label{sfig:gemm_2d}\Ttor{} 2D GEMM. Red = small AMs, blue = large AMs.]{
    \begin{tikzpicture}
      \pgfplotsset{
        xlabel={Cores (16 per node)},xtick={1,8,64},xticklabels={16,128,1024},
        ylabel={Time [sec.]},ymin=0.2,ymax=75,width=4.3cm,
        ylabel near ticks,
      }
      \begin{loglogaxis}[font=\footnotesize]

        \addplot[red, only marks] table[x=nranks,y=total_time] {gemm_data/ttor_2d_nonlarge.dat};
        \addplot[blue,only marks] table[x=nranks,y=total_time] {gemm_data/ttor_2d_large.dat};

        \addplot[red,dashed] table[x=nranks,y=total_time,discard if not={matrix_size}{8192}] {gemm_data/ttor_2d_nonlarge.dat};
        \addplot[red,dashed] table[x=nranks,y=total_time,discard if not={matrix_size}{16384}] {gemm_data/ttor_2d_nonlarge.dat};
        \addplot[red,dashed] table[x=nranks,y=total_time,discard if not={matrix_size}{32768}] {gemm_data/ttor_2d_nonlarge.dat};
        \addplot[red,dashed] table[x=nranks,y=total_time,discard if not={matrix_size}{65536}] {gemm_data/ttor_2d_nonlarge.dat};
        \addplot[black,dotted] table[x=nranks,y=total_time,discard if not={flops_per_rank}{68719476736}] {gemm_data/ttor_2d_nonlarge.dat};
        \addplot[black,dotted] table[x=nranks,y=total_time,discard if not={flops_per_rank}{549755813888}] {gemm_data/ttor_2d_nonlarge.dat};
        \addplot[black,dotted] table[x=nranks,y=total_time,discard if not={flops_per_rank}{4398046511104}] {gemm_data/ttor_2d_nonlarge.dat};

        \addplot[blue,dashed] table[x=nranks,y=total_time,discard if not={matrix_size}{8192}] {gemm_data/ttor_2d_large.dat};
        \addplot[blue,dashed] table[x=nranks,y=total_time,discard if not={matrix_size}{16384}] {gemm_data/ttor_2d_large.dat};
        \addplot[blue,dashed] table[x=nranks,y=total_time,discard if not={matrix_size}{32768}] {gemm_data/ttor_2d_large.dat};
        \addplot[blue,dashed] table[x=nranks,y=total_time,discard if not={matrix_size}{65536}] {gemm_data/ttor_2d_large.dat};
        \addplot[black,dotted] table[x=nranks,y=total_time,discard if not={flops_per_rank}{68719476736}] {gemm_data/ttor_2d_large.dat};
        \addplot[black,dotted] table[x=nranks,y=total_time,discard if not={flops_per_rank}{549755813888}] {gemm_data/ttor_2d_large.dat};
        \addplot[black,dotted] table[x=nranks,y=total_time,discard if not={flops_per_rank}{4398046511104}] {gemm_data/ttor_2d_large.dat};
      
        \node (text) at (axis cs:1,2) {8k};
        \node (text) at (axis cs:1,14) {16k};
        \node (text) at (axis cs:8,18) {32k};
        \node (text) at (axis cs:60,21) {64k};

      \end{loglogaxis}
    \end{tikzpicture}
  }\,
  \subfloat[\label{sfig:gemm_3d}\Ttor{} 3D GEMM. Red = non-tiled, blue = tiled.]{
    \begin{tikzpicture}
      \pgfplotsset{
        xlabel={Cores (16 per node)},xtick={1,8,64},xticklabels={16,128,1024},
        ymin=0.2,ymax=75,width=4.3cm,yticklabels=\empty,
      }
      \begin{loglogaxis}[font=\footnotesize]

        \addplot[red, only marks] table[x=nranks,y=total_time] {gemm_data/ttor_3d_nontiled.dat};
        \addplot[blue,only marks] table[x=nranks,y=total_time] {gemm_data/ttor_3d_tiled.dat};

        \addplot[red,dashed] table[x=nranks,y=total_time,discard if not={matrix_size}{8192}] {gemm_data/ttor_3d_nontiled.dat};
        \addplot[red,dashed] table[x=nranks,y=total_time,discard if not={matrix_size}{16384}] {gemm_data/ttor_3d_nontiled.dat};
        \addplot[red,dashed] table[x=nranks,y=total_time,discard if not={matrix_size}{32768}] {gemm_data/ttor_3d_nontiled.dat};
        \addplot[red,dashed] table[x=nranks,y=total_time,discard if not={matrix_size}{65536}] {gemm_data/ttor_3d_nontiled.dat};
        \addplot[black,dotted] table[x=nranks,y=total_time,discard if not={flops_per_rank}{68719476736}] {gemm_data/ttor_3d_nontiled.dat};
        \addplot[black,dotted] table[x=nranks,y=total_time,discard if not={flops_per_rank}{549755813888}] {gemm_data/ttor_3d_nontiled.dat};
        \addplot[black,dotted] table[x=nranks,y=total_time,discard if not={flops_per_rank}{4398046511104}] {gemm_data/ttor_3d_nontiled.dat};

        \addplot[blue,dashed] table[x=nranks,y=total_time,discard if not={matrix_size}{8192}] {gemm_data/ttor_3d_tiled.dat};
        \addplot[blue,dashed] table[x=nranks,y=total_time,discard if not={matrix_size}{16384}] {gemm_data/ttor_3d_tiled.dat};
        \addplot[blue,dashed] table[x=nranks,y=total_time,discard if not={matrix_size}{32768}] {gemm_data/ttor_3d_tiled.dat};
        \addplot[blue,dashed] table[x=nranks,y=total_time,discard if not={matrix_size}{65536}] {gemm_data/ttor_3d_tiled.dat};
        \addplot[black,dotted] table[x=nranks,y=total_time,discard if not={flops_per_rank}{68719476736}] {gemm_data/ttor_3d_tiled.dat};
        \addplot[black,dotted] table[x=nranks,y=total_time,discard if not={flops_per_rank}{549755813888}] {gemm_data/ttor_3d_tiled.dat};
        \addplot[black,dotted] table[x=nranks,y=total_time,discard if not={flops_per_rank}{4398046511104}] {gemm_data/ttor_3d_tiled.dat};
      
        \node (text) at (axis cs:1,2) {8k};
        \node (text) at (axis cs:1,14) {16k};
        \node (text) at (axis cs:8,18) {32k};
        \node (text) at (axis cs:60,21) {64k};

      \end{loglogaxis}
    \end{tikzpicture}
  } \\
  \subfloat[\label{sfig:gemm_starpu}StarPU 2D GEMM.]{
    \begin{tikzpicture}
      \pgfplotsset{
        xlabel={Cores (16 per node)},xtick={1,8,64},xticklabels={16,128,1024},
        ylabel={Time [sec.]},ymin=0.2,ymax=75,width=4.3cm,
        ylabel near ticks,
      }
      \begin{loglogaxis}[font=\footnotesize]
        \addplot[black,only marks] table[x=nranks,y=total_time] {gemm_data/starpu_2d.dat};
        \addplot[black,dashed] table[x=nranks,y=total_time,discard if not={matrix_size}{8192}] {gemm_data/starpu_2d.dat};
        \addplot[black,dashed] table[x=nranks,y=total_time,discard if not={matrix_size}{16384}] {gemm_data/starpu_2d.dat};
        \addplot[black,dashed] table[x=nranks,y=total_time,discard if not={matrix_size}{32768}] {gemm_data/starpu_2d.dat};
        \addplot[black,dotted] table[x=nranks,y=total_time,discard if not={flops_per_rank}{68719476736}] {gemm_data/starpu_2d.dat};
        \addplot[black,dotted] table[x=nranks,y=total_time,discard if not={flops_per_rank}{549755813888}] {gemm_data/starpu_2d.dat};
        \addplot[black,dotted] table[x=nranks,y=total_time,discard if not={flops_per_rank}{4398046511104}] {gemm_data/starpu_2d.dat};

        \node (text) at (axis cs:1,2) {8k};
        \node (text) at (axis cs:1,14) {16k};
        \node (text) at (axis cs:8,18) {32k};
        \node (text) at (axis cs:60,20) {64k};

      \end{loglogaxis}
    \end{tikzpicture}
  } \,
  \subfloat[\label{sfig:gemm_scalapack}ScaLAPACK GEMM.]{
    \begin{tikzpicture}
      \pgfplotsset{
        xlabel={Cores (16 per node)},xtick={1,8,64},xticklabels={16,128,1024},
        ymin=0.2,ymax=75,width=4.3cm,ylabel near ticks,yticklabels=\empty,
      }
      \begin{loglogaxis}[font=\footnotesize]
        \addplot[black,only marks] table[x=nranks,y=total_time] {gemm_data/scalapack.dat};
        \addplot[black,dashed] table[x=nranks,y=total_time,discard if not={matrix_size}{8192}] {gemm_data/scalapack.dat};
        \addplot[black,dashed] table[x=nranks,y=total_time,discard if not={matrix_size}{16384}] {gemm_data/scalapack.dat};
        \addplot[black,dashed] table[x=nranks,y=total_time,discard if not={matrix_size}{32768}] {gemm_data/scalapack.dat};
        \addplot[black,dotted] table[x=nranks,y=total_time,discard if not={flops_per_rank}{68719476736}] {gemm_data/scalapack.dat};
        \addplot[black,dotted] table[x=nranks,y=total_time,discard if not={flops_per_rank}{549755813888}] {gemm_data/scalapack.dat};
        \addplot[black,dotted] table[x=nranks,y=total_time,discard if not={flops_per_rank}{4398046511104}] {gemm_data/scalapack.dat};

        \node (text) at (axis cs:1,2) {8k};
        \node (text) at (axis cs:1,18) {16k};
        \node (text) at (axis cs:8,30) {32k};
        \node (text) at (axis cs:60,27) {64k};

      \end{loglogaxis}
    \end{tikzpicture}
  } \\
  \subfloat[\label{sfig:gemm_blocksize}Block size impact, $N = 32\,768$ with 1024 CPUs.]{
    \begin{tikzpicture}
      \pgfplotsset{
        xlabel={Block size},xtick={64,256,1024},xticklabels={64,256,1024},
        width=4.3cm,ylabel near ticks,ylabel={Time [sec.]},
      }
      \begin{loglogaxis}[font=\footnotesize,  legend style={nodes={scale=0.7, transform shape}}]
        \addplot[black,mark=*] table[x=block_size,y=total_time] {gemm_data/ttor_2d_large_blocksize.dat};
        \addplot[black,dashed,mark=triangle*,mark options={solid}] table[x=block_size,y=total_time] {gemm_data/starpu_2d_nonpruned_blocksize.dat};
        \addplot[black,mark=square*] table[x=block_size,y=total_time,discard if not={block_size}{64}] {gemm_data/ttor_2d_nonlarge_blocksize.dat};        
        \legend{\Ttor{} (2D), StarPU};
      \end{loglogaxis}
      \node (text) at (0.3,0.42) [fill=white,inner sep=1pt,anchor=north,align=left,font=\scriptsize] {Small AMs};
      \draw [->,>=stealth] (text)--(0.24,0.9);
    \end{tikzpicture}
  } \,
    \subfloat[\label{sfig:gemm_concurrency} \Ttor{} 2D GEMM. Concurrency = \lstinline{num_blocks^2/n_cores},  $N = 16\,384$.]{
    \begin{tikzpicture}
      \pgfplotsset{
	      xlabel={Concurrency}, log basis x = {2}, xtick={1,16,256, 4096}, xticklabels={1,16,256, 4096},
	      ymin=0,ymax=1,yticklabel={\pgfmathparse{\tick*100}\pgfmathprintnumber{\pgfmathresult}\%}, ylabel={Efficiency}, ylabel near ticks,
        width=4.3cm
      }
	    \begin{axis}[font=\footnotesize, xmode = log, legend style={draw=none, fill=none, at={(1.0,0.79)},  nodes={scale=0.6, transform shape}}]
	      \addlegendimage{empty legend}
	      \addlegendentry{\# nodes}
	      \addplot[red, mark=x, thick,only marks] table[x=conc,y=eff, discard if not={n}{16384}, discard if not={nodes}{1}] {gemm_data/ttor_eff_conc.dat};
	      \addlegendentry{1};
	      \addplot[orange, mark=o, thick,only marks] table[x=conc,y=eff, discard if not={n}{16384}, discard if not={nodes}{2}] {gemm_data/ttor_eff_conc.dat};
	      \addlegendentry{2};  
	      \addplot[teal, mark=square, thick,only marks] table[x=conc,y=eff, discard if not={n}{16384}, discard if not={nodes}{4}] {gemm_data/ttor_eff_conc.dat};
	      \addlegendentry{4};  
	      \addplot[blue, mark=+, thick,only marks] table[x=conc,y=eff, discard if not={n}{16384}, discard if not={nodes}{8}] {gemm_data/ttor_eff_conc.dat};
	      \addlegendentry{8};  
	      \addplot[purple, mark=triangle, thick,only marks] table[x=conc,y=eff, discard if not={n}{16384}, discard if not={nodes}{16}] {gemm_data/ttor_eff_conc.dat};
              \addlegendentry{16};  
              \addplot[green!40!gray, mark=o, thick,only marks] table[x=conc,y=eff, discard if not={n}{16384}, discard if not={nodes}{32}] {gemm_data/ttor_eff_conc.dat};
              \addlegendentry{32}; 
      \end{axis}
    \end{tikzpicture}
  } 
  
  \caption{GEMM scalings.
    (a-b): impact of task granularity on 3D GEMM. Smaller tasks give higher overlap of computation and communication.
    (c-f): weak (dotted) and strong (dashed) scalings. Numbers indicate the matrix size $N$. Largest test case is $N = 65\,536$. 
    (g): optimal block size (i.e., task granularity) for the $N=32\,768$ test case.
    The extra data point shows the improvement when using small AMs instead of large AMs on small block sizes. The decrease in the number of messages sent improves the runtime by 3x.
    (h): efficiency as a function of concurrency for $N=16\,384$. Reference timing is with 1 core.}
  \label{fig:gemm_scaling}
\end{figure}

%% file: dense_cholesky.tex
\subsection{Distributed dense cholesky factorization}

We now consider an implementation of the Cholesky algorithm, i.e., given a symmetric positive definite matrix $A \in \mathbb{R}^{N \times N}$, compute $L$ such that
$A = L L^\top$.
In its sequential and blocked form, the algorithm is described in \Cref{alg:cholesky_sequential}.

\begin{algorithm}
  \begin{algorithmic}[1]
    \Procedure{Cholesky}{$A$, $n$} \Comment{$A \succ 0$, $n \times n$ blocks}
      \For{$1 \leq k \leq n$}
        \State $L_{kk} L_{kk}^\top = A_{kk}$ \Comment{potrf($k$)}
        \For{$k+1 \leq i \leq n$}
          \State $L_{ik} = A_{ik} L_{kk}^{-\top}$ \Comment{trsm($i,k$)}
          \For{$k+1 \leq j \leq i$}
            \State $A_{ij} \leftarrow A_{ij} - L_{ik} L_{jk}^\top$ \Comment{gemm($k,i,j$)}
          \EndFor
        \EndFor
      \EndFor
    \EndProcedure
  \end{algorithmic}
  \caption{}
  \label{alg:cholesky_sequential}
\end{algorithm}

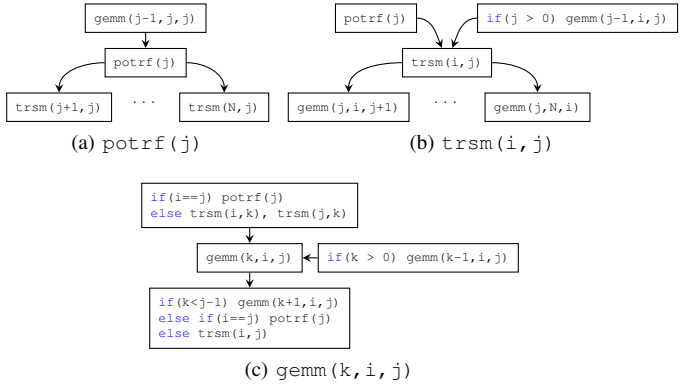
\begin{figure}
  \centering
  \subfloat[][\lstinline{potrf(j)}]{
    \begin{tikzpicture}
      \tikzstyle{every node}=[rectangle,draw,font=\tiny,text centered];
      \def\L{0.2cm}; 
      \node (potrf_j) {\lstinline{potrf(j)}};
      \node [above = \L of potrf_j] (a) {\lstinline{gemm(j-1,j,j)}};
      \node [below left = \L and -0.1cm of potrf_j] (b) {\lstinline{trsm(j+1,j)}};
      \node [below right = \L and -0.1cm of potrf_j] (c) {\lstinline{trsm(N,j)}};
      \path [->,>=stealth] (a) edge (potrf_j);
      \path [->,>=stealth] (potrf_j) edge[out=west,in=north] (b);
      \path [->,>=stealth] (potrf_j) edge[out=east,in=north] (c);
      \tikzstyle{every node}=[font=\tiny,text centered];
      \node [below = \L of potrf_j] {$\dots$};
    \end{tikzpicture}
  }
  \subfloat[][\lstinline{trsm(i,j)}]{
    \begin{tikzpicture}
      \tikzstyle{every node}=[rectangle,draw,font=\tiny,text centered];
      \def\L{0.2cm};
      \node (trsm_ij) {\lstinline{trsm(i,j)}};
      \node [above left = \L and -0.2cm of trsm_ij] (a) {\lstinline{potrf(j)}};
      \node [above right = \L and -0.2cm of trsm_ij] (b) {\lstinline{if(j > 0) gemm(j-1,i,j)}};
      \node [below left = \L and -0.1cm of trsm_ij] (c) {\lstinline{gemm(j,i,j+1)}};
      \node [below right = \L and -0.1cm of trsm_ij] (d) {\lstinline{gemm(j,N,i)}};
      \path [->,>=stealth] (a) edge[out=east,in=110] (trsm_ij);
      \path [->,>=stealth] (b) edge[out=west,in=70] (trsm_ij);
      \path [->,>=stealth] (trsm_ij) edge[out=west,in=north] (c);
      \path [->,>=stealth] (trsm_ij) edge[out=east,in=north] (d);
      \tikzstyle{every node}=[font=\tiny,text centered];
      \node [below = \L of trsm_ij] {$\dots$};
    \end{tikzpicture}
  } \\
  \subfloat[][\lstinline{gemm(k,i,j)}]{
    \begin{tikzpicture}
      \tikzstyle{every node}=[rectangle,draw,font=\tiny,text centered];
      \def\L{0.2cm};
      \node (gemm_kij) {\lstinline{gemm(k,i,j)}};
      \node [above = \L of gemm_kij,align=left] (a)
      {\lstinline{if(i==j) potrf(j)}\\
      \lstinline{else trsm(i,k), trsm(j,k)}};
      \node [right = \L of gemm_kij] (b) {\lstinline{if(k > 0) gemm(k-1,i,j)}};
      \node [below = \L of gemm_kij,align=left] (c)
      {\lstinline{if(k<j-1) gemm(k+1,i,j)}\\
      \lstinline{else if(i==j) potrf(j)}\\
      \lstinline{else trsm(i,j)}};
      \path [->,>=stealth] (a) edge[out=south,in=north] (gemm_kij);
      \path [->,>=stealth] (b) edge[out=west,in=east] (gemm_kij);
      \path [->,>=stealth] (gemm_kij) edge (c);
    \end{tikzpicture}
  }
  \caption{PTG description of \Cref{alg:cholesky_sequential}. 
  In \Ttor{}, when out-dependencies are remote, an AM is sent to the remote rank, carrying the associated block and triggering remote tasks.}
  \label{fig:cholesky_ptg}
\end{figure}

The algorithm is made of three main computational routines: potrf($k$), trsm($i,k$) and gemm($k,i,j$) (in practice syrk when $i = j$). We show a PTG formulation of \Cref{alg:cholesky_sequential} in \Cref{fig:cholesky_ptg}. Large active messages are used.

We compare \Ttor{}, StarPU (with STF semantics) and ScaLAPACK. A 2D block cyclic data distribution is used with a block size of 256. Task priorities in \Ttor{} are computed using~\cite{beaumont2020makespan}. As before, in ScaLAPACK the block size is related to the data distribution but there are no tasks per se.

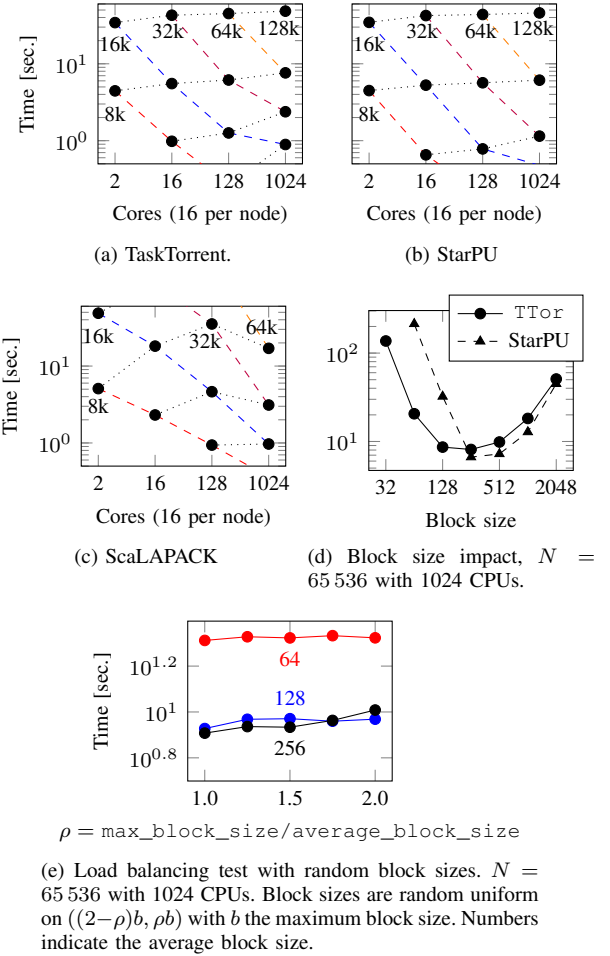
\begin{figure}
  \centering
  \subfloat[TaskTorrent.]{
    \begin{tikzpicture}
      \begin{loglogaxis}[
          xlabel={Cores (16 per node)},xtick={2,16,128,1024},xticklabels={2,16,128,1024},
          ymin=0.5,ymax=60,ylabel={Time [sec.]},width=4.3cm,font=\footnotesize,
          ylabel near ticks,xlabel near ticks,
        ]
        \addplot[black,only marks] table[x=total_ncores,y=total_time] {dense_cholesky_leo/ttor.dat};
        \addplot[black,dotted] table[x=total_ncores,y=total_time,discard if not={flops_per_core}{536870912}] {dense_cholesky_leo/ttor.dat};
        \addplot[black,dotted] table[x=total_ncores,y=total_time,discard if not={flops_per_core}{4294967296}] {dense_cholesky_leo/ttor.dat};
        \addplot[black,dotted] table[x=total_ncores,y=total_time,discard if not={flops_per_core}{34359738368}] {dense_cholesky_leo/ttor.dat};
        \addplot[black,dotted] table[x=total_ncores,y=total_time,discard if not={flops_per_core}{274877906944}] {dense_cholesky_leo/ttor.dat};
        \addplot[black,dotted] table[x=total_ncores,y=total_time,discard if not={flops_per_core}{2199023255552}] {dense_cholesky_leo/ttor.dat};
        \addplot[red,dashed] table[x=total_ncores,y=total_time,discard if not={matrix_size}{8192}] {dense_cholesky_leo/ttor.dat};
        \addplot[blue,dashed] table[x=total_ncores,y=total_time,discard if not={matrix_size}{16384}] {dense_cholesky_leo/ttor.dat};
        \addplot[purple,dashed] table[x=total_ncores,y=total_time,discard if not={matrix_size}{32768}] {dense_cholesky_leo/ttor.dat};
        \addplot[orange,dashed] table[x=total_ncores,y=total_time,discard if not={matrix_size}{65536}] {dense_cholesky_leo/ttor.dat};
        \addplot[black,dashed] table[x=total_ncores,y=total_time,discard if not={matrix_size}{131072}] {dense_cholesky_leo/ttor.dat};
        \node (text) at (axis cs:2,2.2) {8k};
        \node (text) at (axis cs:2,18) {16k};
        \node (text) at (axis cs:14,27) {32k};
        \node (text) at (axis cs:120,27) {64k};
        \node (text) at (axis cs:800,29) {128k};
      \end{loglogaxis}
    \end{tikzpicture}
  }
  \subfloat[StarPU]{
    \begin{tikzpicture}
      \begin{loglogaxis}[
          xlabel={Cores (16 per node)},xtick={2,16,128,1024},xticklabels={2,16,128,1024},
          ymin=0.5,ymax=60,yticklabels=\empty,width=4.3cm,font=\footnotesize,
          ylabel near ticks,xlabel near ticks,
        ]
        \addplot[black,only marks] table[x=total_ncores,y=total_time] {dense_cholesky_leo/starpu.dat};
        \addplot[black,dotted] table[x=total_ncores,y=total_time,discard if not={flops_per_core}{536870912}] {dense_cholesky_leo/starpu.dat};
        \addplot[black,dotted] table[x=total_ncores,y=total_time,discard if not={flops_per_core}{4294967296}] {dense_cholesky_leo/starpu.dat};
        \addplot[black,dotted] table[x=total_ncores,y=total_time,discard if not={flops_per_core}{34359738368}] {dense_cholesky_leo/starpu.dat};
        \addplot[black,dotted] table[x=total_ncores,y=total_time,discard if not={flops_per_core}{274877906944}] {dense_cholesky_leo/starpu.dat};
        \addplot[black,dotted] table[x=total_ncores,y=total_time,discard if not={flops_per_core}{2199023255552}] {dense_cholesky_leo/starpu.dat};
        \addplot[red,dashed] table[x=total_ncores,y=total_time,discard if not={matrix_size}{8192}] {dense_cholesky_leo/starpu.dat};
        \addplot[blue,dashed] table[x=total_ncores,y=total_time,discard if not={matrix_size}{16384}] {dense_cholesky_leo/starpu.dat};
        \addplot[purple,dashed] table[x=total_ncores,y=total_time,discard if not={matrix_size}{32768}] {dense_cholesky_leo/starpu.dat};
        \addplot[orange,dashed] table[x=total_ncores,y=total_time,discard if not={matrix_size}{65536}] {dense_cholesky_leo/starpu.dat};
        \addplot[black,dashed] table[x=total_ncores,y=total_time,discard if not={matrix_size}{131072}] {dense_cholesky_leo/starpu.dat};
        \node (text) at (axis cs:2,2.2) {8k};
        \node (text) at (axis cs:2,18) {16k};
        \node (text) at (axis cs:14,27) {32k};
        \node (text) at (axis cs:120,27) {64k};
        \node (text) at (axis cs:800,29) {128k};
      \end{loglogaxis}
    \end{tikzpicture}
  } \\
  \subfloat[ScaLAPACK]{
    \begin{tikzpicture}
      \begin{loglogaxis}[
        xlabel={Cores (16 per node)},xtick={2,16,128,1024},xticklabels={2,16,128,1024},
        ymin=0.5,ymax=60,width=4.3cm,font=\footnotesize,ylabel={Time [sec.]},
        ylabel near ticks,xlabel near ticks,
        ]
        \addplot[black,only marks] table[x=total_ncores,y=total_time] {dense_cholesky_leo/scalapack.dat};
        \addplot[black,dotted] table[x=total_ncores,y=total_time,discard if not={flops_per_core}{536870912}] {dense_cholesky_leo/scalapack.dat};
        \addplot[black,dotted] table[x=total_ncores,y=total_time,discard if not={flops_per_core}{4294967296}] {dense_cholesky_leo/scalapack.dat};
        \addplot[black,dotted] table[x=total_ncores,y=total_time,discard if not={flops_per_core}{34359738368}] {dense_cholesky_leo/scalapack.dat};
        \addplot[black,dotted] table[x=total_ncores,y=total_time,discard if not={flops_per_core}{274877906944}] {dense_cholesky_leo/scalapack.dat};
        \addplot[black,dotted] table[x=total_ncores,y=total_time,discard if not={flops_per_core}{2199023255552}] {dense_cholesky_leo/scalapack.dat};
        \addplot[red,dashed] table[x=total_ncores,y=total_time,discard if not={matrix_size}{8192}] {dense_cholesky_leo/scalapack.dat};
        \addplot[blue,dashed] table[x=total_ncores,y=total_time,discard if not={matrix_size}{16384}] {dense_cholesky_leo/scalapack.dat};
        \addplot[purple,dashed] table[x=total_ncores,y=total_time,discard if not={matrix_size}{32768}] {dense_cholesky_leo/scalapack.dat};
        \addplot[orange,dashed] table[x=total_ncores,y=total_time,discard if not={matrix_size}{65536}] {dense_cholesky_leo/scalapack.dat};
        \addplot[black,dashed] table[x=total_ncores,y=total_time,discard if not={matrix_size}{131072}] {dense_cholesky_leo/scalapack.dat};
        \node (text) at (axis cs:2,3) {8k};
        \node (text) at (axis cs:2,25) {16k};
        \node (text) at (axis cs:100,21) {32k};
        \node (text) at (axis cs:800,32) {64k};
      \end{loglogaxis}
    \end{tikzpicture}
  }
  \subfloat[Block size impact, $N = 65\,536$ with 1024 CPUs.]{
    \label{sfig:cholesky_scalings_large}
    \begin{tikzpicture}
      \begin{loglogaxis}[
          font=\footnotesize,xlabel={Block size},xtick={32,128,512,2048},
          xticklabels={32,128,512,2048},width=4.3cm,
          ylabel near ticks, xlabel near ticks,
          legend style={at={(1.05,1.1)}}
         ]
        \addplot[mark=*]
          table[x=block_size,y=total_time] {dense_cholesky_leo/ttor_blocksize.dat};
        \addplot[dashed,mark=triangle*,mark options={solid}] 
          table[x=block_size,y=total_time] {dense_cholesky_leo/starpu_blocksize_nopruning.dat};
        \legend{\Ttor{}, StarPU};
      \end{loglogaxis}
    \end{tikzpicture}
  } \\
  \subfloat[Load balancing test with random block sizes. $N = 65\,536$ with 1024 CPUs.
  Block sizes are random uniform on $((2-\rho)b,\rho b)$ with $b$ the maximum block size.
  Numbers indicate the average block size.]{
    \label{sfig:cholesky_random_block}
    \begin{tikzpicture}
      \begin{semilogyaxis}[
          font=\footnotesize,xtick={1,1.5,2.0},
          xticklabels={1.0,1.5,2.0},width=4.3cm,
          ylabel near ticks,xlabel near ticks,ylabel={Time [sec.]},
          ymin=5,xlabel={$\rho = $ \lstinline{max_block_size/average_block_size}},
         ]
        \addplot[red,mark=*] table[x=ratio,y=total_time,discard if not={block_size}{64}]  {dense_cholesky_leo/ttor_random.dat};
        \addplot[blue,mark=*] table[x=ratio,y=total_time,discard if not={block_size}{128}] {dense_cholesky_leo/ttor_random.dat};
        \addplot[mark=*] table[x=ratio,y=total_time,discard if not={block_size}{256}] {dense_cholesky_leo/ttor_random.dat};
        \node at (axis cs:1.5,7)    [align=left] {\textcolor{black}{256}};
        \node at (axis cs:1.5,11.5)   [align=left] {\textcolor{blue}{128}};
        \node at (axis cs:1.5,17)   [align=left] {\textcolor{red}{64}};
      \end{semilogyaxis}
    \end{tikzpicture}
  }
  \caption{Cholesky scalings.
  (a-c): weak (dotted) and strong (dashed) scalings. Numbers indicate the matrix size $N$. Largest (top right) test case is $N = 131\,072$.
  (d): optimal block size (i.e., task granularity) for the $N=65\,536$ test case.
  (e): load balancing test using random block sizes for the $N=65\,536$ test case.
  }
  \label{fig:cholesky_scalings}
\end{figure}

Weak and strong scalings are performed by multiplying the number of rows and columns by 2 or the number of cores by 8. The larger test case is a matrix of size $N = 131\,072$. \Cref{fig:cholesky_scalings} shows the results.

We see that on large problems, both \Ttor{} and StarPU reach very similar performances, both outperforming ScaLAPACK by far: for $N = 131\,072$ on 1024 cores, ScaLAPACK takes more than $125$ secs (not shown). On the $N = 131\,072$ test case, \Ttor{} and StarPU differ by less than 10\%. \added{StarPU shows better strong scaling for small problems on many nodes. We conjecture that this may be due to a better task scheduler, memory management (thread-memory affinity), and mapping of the computation across nodes.}

\Cref{sfig:cholesky_scalings_large} shows the runtime as a function of the block size for a test case of size $65\,536 \times 65\,536$ on 64 nodes (1024 CPUs). We see that 256 gives the best results for both \Ttor{} and StarPU. Furthermore, we observe that for small task size, \Ttor{} degrades less quickly than StarPU. The small block size leads to many tasks and unrolling the DAG on one node becomes prohibitive, even for reasonably large tasks (block size of 128). For a block size of 64, \Ttor{} is about 10x faster. Thanks to its lightweight runtime and distributed DAG exploration, \Ttor{} suffers less from the small task size. For large task sizes, both degrade similarly. The poor performance at large size is caused by a lack of concurrency.

\Cref{sfig:cholesky_random_block} shows a load balancing test using random block sizes with a fixed number of blocks. $\rho$ is the ratio of the largest over the average block size. For $\rho = 1.5$, the ratio of flops from smallest to largest task is $(1.5/0.5)^3 = 27$. We see that \Ttor{} handles tasks of various granularity very well, with less than 25\% degradation from $\rho = 1$ to $\rho = 2$ for an average block size of 256.

%% file: previous.tex
\section{Previous work}
\label{sec:previous_work}

\paragraph{Runtime systems}

As mentioned in \Cref{subsec:stf_ptg}, other task-based runtime systems exist. We highlight some of their characteristics.
PaRSEC \cite{6654146} is a runtime system centered around dense linear algebra. It takes the PTG approach but uses a custom programming language, the JDF. 
This can make adoption harder for new users.
Legion \cite{bauer2014legion} is a general purpose STF runtime. 
It has many features and can be used from C++ but requires the user to express everything using Legion's data structures.
It is also intended to be used primarily with GASNet \cite{bonachea2017gasnet} and not MPI.
Regent \cite{slaughter2015regent} proposes a higher level language on top of Legion, making programming more productive. 
Unfortunately, obtaining high performance requires the user to program directly the mapper 
which is time-consuming and requires a detailed understanding of the inner workings of Legion.
Finally, StarPU \cite{augonnet2011starpu} uses C++ and is STF-based.
The data is initially distributed by the user like a classical MPI code,
and various scheduling strategies can be used to further improve performance.
However, user data still has to be wrapped using StarPU's data structures.

In designing \Ttor{} we chose to focus on the following features.
The message passing paradigm requires the programmer to distribute data
but simplifies the design of the library with the goal of minimizing global synchronization and communication.
MPI and C++ makes integration into other codes easier.
Active messages are necessary because of the asynchronous nature of computations.
Finally the PTG approach leads to a minimal runtime overhead. 
Note however that the choice of PTG has drawbacks: it can be difficult for the programmer to reason about tasks dependencies.
This can be easier in some applications (like linear algebra) than others.
\Ttor{} also does not consider concepts like memory affinity or accelerators at the moment. This is reserved for future work.

\paragraph{Task-based parallelism}
Task-based parallelism is now a common feature of many parallel programming systems.

Cilk \cite{joerg1996cilk, frigo1998implementation} introduced a multi-threading component to C in 1996, and Cilk-5 introduced \lstinline{spawn} and asynchronous computations.
Many other efforts followed, including OpenMP \cite{dagum1998openmp} (with tasking introduced in version 3.0), 
Intel TBB \cite{reinders2007intel} (where task DAGs can be expressed), Cilk Plus \cite{robison2013composable}, XKaapi \cite{gautier2013xkaapi},
OmpSs \cite{duran2011ompss}, Superglue \cite{tillenius2015superglue}, and the SMPSs programming model \cite{perez2008dependency, perez2010handling}.
The Plasma \cite{agullo2009numerical, agullo2009plasma} (for CPU) and Magma \cite{tomov2009magma} (for CPU and GPU) libraries
are replacements for multithreaded LAPACK, where parallelism is obtained through tiled algorithms using a dynamic runtime, Quark \cite{yarkhan2011quark}.

Notice that all the previously mentioned work is typically only usable in a shared-memory context.
In particular, there is no support to let one rank trigger (or fulfill the dependency of) a task on another rank.

\paragraph{Distributed programming}
An explicit goal of \Ttor{} is to provide support for distributed computing.

The most common distributed programming paradigm is using explicit message passing like in MPI. 
In MPI, ranks are completely independent and only communicate with each other through explicit message passing. 
Charm++ \cite{kale1993charm++} takes an object-oriented approach. It exposes \emph{chares} which are concurrent objects communicating through messages.
We also mention \lstinline{DARMA/vt} \cite{lifflander2020darma}, a tasking and active message library in C++, 
with other features such as load balancing and asynchronous collectives.
Finally, in the PGAS (partitioned global address space) model (like GASNet \cite{bonachea2017gasnet}), each rank can access a global address space through read (get) and write (put) operations.
Chapel \cite{callahan2004cascade}, Fortran Co-arrays \cite{numrich1998co}, UPC \cite{el2006upc} and UPC++ \cite{zheng2014upcxx} are examples of PGAS-based parallel programming languages.

\paragraph{Active messages}
One-sided active messages is another important feature of TaskTorrent. 
Von Eicken et al. \cite{von1992active} argued in 1992 that active messages are a powerful mechanism to hide latency and improve performance. 
Active messages are also a central part of UPC++ where they resemble the ones in \Ttor{}. 
In UPC++, however, remote data is referred to using global data structures, 
while \Ttor{} tends to use the C++ variable capture mechanism in lambda functions. 




%% file: conclusion.tex
\section{Conclusion}

We presented TaskTorrent (\Ttor{}), a lightweight distributed task-based runtime system in C++. It has a friendly API, and rely on readily available tools (C++14 and MPI). It enables shared-memory task-based parallelism coupled with one-sided active messages. Those two concepts naturally work together to create a distributed task-based parallel computing framework. We showed that \Ttor{} is competitive with both StarPU (a state of the art runtime) and ScaLAPACK on large problems. Its lightweight nature allows it to be more forgiving when task granularity is not optimal, which is key to integrating this approach in legacy codes.

%% file: appendix.tex
\input{ad_ae.tex}

%% file: ad_ae.tex
\section{AD-AE Appendix}

\subsection{Paper Artifact Description / Article Evaluation (AD/AE) Appendix}

Are there computational artifacts such as datasets, software, or hardware associated with this paper? Yes.

\subsection{AD/AE Details}

Experiments were run on a Stanford University HPC cluster equipped with dual-sockets and 16 cores \lstinline[breaklines=true]{Intel(R) Xeon(R) CPU E5-2670 0 @ 2.60GHz} with 32GB of RAM per node. 
Intel Compiler (icpc (ICC)  \lstinline[breaklines=true]{19.1.0.166 20191121}) and Intel MPI are used with Intel MKL (version \lstinline[breaklines=true]{2020.0.166}) for BLAS, LAPACK and ScaLAPACK. We use StarPU version  \lstinline[breaklines=true]{1.3.2}.

\subsubsection{Artifacts Available (AA)}

\begin{itemize}
    \item Software Artifact Availability: All author-created software artifacts are maintained in a public repository under an OSI-approved license.
    \item Hardware Artifact Availability: There are no author-created hardware artifacts.
    \item Data Artifact Availability: There are no author-created data artifacts.
    \item Proprietary Artifacts: There are associated proprietary artifacts that are not created by the authors. Some author-created artifacts are proprietary.
\end{itemize}

\subsubsection{Author artifacts}

\begin{itemize}
    \item Artifact 1: TaskTorrent repository, \lstinline[breaklines=true]{github.com/leopoldcambier/tasktorrent}
    \item Artifact 2: TaskTorrent paper Scalapack and StarPU benchmarks repository, \lstinline[breaklines=true]{github.com/leopoldcambier/tasktorrent_paper_benchmarks}
\end{itemize}

\subsubsection{Experimental setup}

\begin{itemize}
    \item Relevant hardware: Dual socket \lstinline[breaklines=true]{Intel(R) Xeon(R) CPU E5-2670 0 @ 2.60GHz} with 16 cores and 32GB of RAM per node.
    \item Operating systems and versions: Linux kernel \lstinline{3.10.0-693.el7.x86_64}
    \item Compilers and versions: Intel Compiler (icpc (ICC) \lstinline[breaklines=true]{19.1.0.166 20191121})
    \item Applications and versions: N/A
    \item Libraries and versions: Intel MPI version \lstinline{2018.2.199}, Intel MKL version \lstinline{19.1.0.166}, StarPU version \lstinline{1.3.2}
    \item Key algorithms: N/A
    \item Input datasets and versions: N/A
    \item Optional link (URL) to output from commands that gather execution environment information: \lstinline[breaklines=true]{stanford.edu/~lcambier/tasktorrent_paper/AD_AE.txt}
\end{itemize}

\subsection{Artifact Evaluation}

Are you completing an Artifact Evaluation (AE) Appendix? No.